\documentclass[%
twocolumn,
 reprint,
 amsmath,amssymb,
 aps,
 superscriptaddress,
 longbibliography
]{revtex4-2}

\usepackage{graphicx}
\usepackage{dcolumn}
\usepackage[dvipsnames]{xcolor}
\usepackage[colorlinks,linkcolor=blue,citecolor=blue,urlcolor=blue]{hyperref}
\usepackage{stix}

\renewcommand{\vec}[1]{\mathbf{#1}}

\begin{document}

\title{Unveiling the Orbital Texture of 1T-TiTe$_2$ using Intrinsic Linear Dichroism in Multidimensional Photoemission Spectroscopy}  

\author{Samuel Beaulieu}
\email{samuel.beaulieu@u-bordeaux.fr}
\affiliation{Universit\'e de Bordeaux - CNRS - CEA, CELIA, UMR5107, F33405 Talence, France}
\affiliation{Fritz Haber Institute of the Max Planck Society, Faradayweg 4-6, 14195 Berlin, Germany}

\author{Michael Sch\"uler}
\email{schuelem@stanford.edu}
\affiliation{Stanford Institute for Materials and Energy Sciences (SIMES),
  SLAC National Accelerator Laboratory, Menlo Park, CA 94025, USA}

\author{Jakub Schusser}
\affiliation{Experimentelle Physik VII and Würzburg-Dresden Cluster of Excellence ct.qmat, Universität Würzburg, Würzburg, Germany}
\affiliation{New Technologies-Research Center, University of West Bohemia, 30614 Pilsen, Czech Republic}

\author{Shuo Dong}
\affiliation{Fritz Haber Institute of the Max Planck Society, Faradayweg 4-6, 14195 Berlin, Germany}

\author{Tommaso Pincelli}
\affiliation{Fritz Haber Institute of the Max Planck Society, Faradayweg 4-6, 14195 Berlin, Germany}

\author{Julian Maklar}
\affiliation{Fritz Haber Institute of the Max Planck Society, Faradayweg 4-6, 14195 Berlin, Germany}

\author{Alexander Neef}
\affiliation{Fritz Haber Institute of the Max Planck Society, Faradayweg 4-6, 14195 Berlin, Germany}

\author{Friedrich Reinert}
\affiliation{Experimentelle Physik VII and Würzburg-Dresden Cluster of Excellence ct.qmat, Universität Würzburg, Würzburg, Germany}

\author{Martin Wolf}
\affiliation{Fritz Haber Institute of the Max Planck Society, Faradayweg 4-6, 14195 Berlin, Germany}

\author{Laurenz Rettig}
\affiliation{Fritz Haber Institute of the Max Planck Society, Faradayweg 4-6, 14195 Berlin, Germany}

\author{J\'an Min\'ar}
\email{jminar@ntc.zcu.cz}
\affiliation{New Technologies-Research Center, University of West Bohemia, 30614 Pilsen, Czech Republic}

\author{Ralph Ernstorfer}
\email{ernstorfer@fhi-berlin.mpg.de}
\affiliation{Fritz Haber Institute of the Max Planck Society, Faradayweg 4-6, 14195 Berlin, Germany}
\affiliation{Institut für Optik und Atomare Physik, Technische Universität Berlin, 10623 Berlin, Germany}

\begin{abstract}

The momentum-dependent orbital character in crystalline solids, referred to as orbital texture, is of capital importance in the emergence of symmetry-broken collective phases such as charge density waves as well as superconducting and topological states of matter. By performing extreme ultraviolet multidimensional angle-resolved photoemission spectroscopy for two different crystal orientations linked to each other by mirror symmetry, we isolate and identify the role of orbital texture in photoemission from the transition metal dichalcogenide 1T-TiTe$_2$. By comparing our experimental results with theoretical calculations based on both a quantitative one-step model of photoemission and an intuitive tight-binding model, we unambiguously demonstrate the link between the momentum-dependent orbital orientation and the emergence of strong intrinsic linear dichroism in the photoelectron angular distributions. Our results represent an important step towards going beyond band structure (eigenvalues) mapping and learn about electronic wavefunction and orbital texture of solids by exploiting matrix element effects in photoemission spectroscopy.    

\end{abstract}

\date{\today}
\maketitle

\section{Introduction}\label{intro}

Angle-resolved photoemission spectroscopy (ARPES) is the most direct experimental technique to measure the momentum-resolved electronic eigenvalues (energy bands) of crystalline solids~\cite{Damascelli04,Sobota21}. The technique is based on the photoelectric effect~\cite{Hertz87,Einstein05}, in which electrons inside solids absorb a photon with energy larger than the work function and escape into the vacuum. A key aspect of ARPES is that the energy ($E$) and momentum ($\textbf{k}$)-resolved photoemission signal is proportional to the single-particle spectral function $A(\textbf{k},E)$~\cite{Hufner99}, a fundamental quantity containing key information about many-body interactions inside solids. 

However, more than two decades ago, it was realized that the direct relationship between ARPES spectral lineshapes and the spectral function was obstructed by additional intensity modulation coming from the transition dipole matrix element connecting the initial Bloch wave and the final photoelectron states~\cite{Bansil99}. Because of its semimetallic 2D nature and its speculated textbook Fermi-Liquid behavior, the layered transition metal dichalcogenide (TMDC) 1T-TiTe$_2$ was the central material around the debate on directly linking the spectral function and the ARPES lineshape~\cite{Claessen92,Claessen96,Allen94,perfetti01}. Indeed, 1T-TiTe$_2$ served as a benchmark material to demonstrate the signatures of many-body effects in ARPES lineshape~\cite{nicolay06}, the possibility to retrieve the photohole lifetime~\cite{Krasovskii07}, as well as to the potential to dissect the role of the different quasiparticle scattering processes from ARPES measurements~\cite{perfetti01}. 
On the other hand, very little attention has been devoted to the other key ingredient governing ARPES spectra of 1T-TiTe$_2$: the transition dipole matrix element. The effect of the transition dipole matrix element is often viewed as detrimental, since it breaks the quasi-one-to-one correspondence between the ARPES spectrum and the occupied electronic band structure, by imprinting a complex and often strong energy- and momentum-dependent modulation onto the photoemission intensity~\cite{Moser17,Day19}. Matrix element effects can sometimes be confused with signatures of many-body interactions~\cite{Bansil99,Inosov07} in ARPES spectra, and are often interpreted by hand-waving arguments. While computational schemes to remove matrix element effects from measured ARPES spectra are an attractive route to focus on the band structure~\cite{xian20}, it has become clear that additional information on the electronic wavefunction can be obtained by understanding matrix element effects. While the matrix element has some extrinsic contributions coming from the experimental geometry (\textit{e.g.} photon momentum, light polarization, and crystal orientation), it also encodes intrinsic information on the nature of the initial state from which the electrons are emitted. Therefore, disentangling extrinsic and intrinsic contributions is of capital importance to access the additional information encoded in the photoemission matrix elements. Important examples include gaining insights into the orbital character of electronic band structure in solids~\cite{Wang12,Park12,Sterzi18,Louat19,bentmann_profiling_2021}, chirality of charge carriers~\cite{Liu11,Bao21}, quantum geometric properties such as orbital pseudospin texture~\cite{Beaulieu20_2,Schuler21} as well as Berry curvature~\cite{Cho18,Schuler20,Cho21,unzelmann_momentum-space_2021}. 
The orbital texture in TMDCs is very sensitive to the lattice structure and collective effects such as charge density waves (CDW)~\cite{Terashima03,Watson19,Strokov12,Chen17,Lin20} and the debated excitonic insulator state~\cite{mor_ultrafast_2017,mazza_nature_2020,baldini_spontaneous_2020}. As demonstrated recently, CDW phases and orbital texture can be intertwined~\cite{peng_observation_2021}. Tracing the orbital texture from matrix element effects in photoemission will also greatly benefit discerning structural and collective phases in TMDCs. 

To disentangle between extrinsic and intrinsic contributions to dichroism in ARPES, we recently introduced a new measurement methodology to extract the modulation of the photoemission intensity upon a crystal rotation by a well-chosen angle $\theta$ in the laboratory frame ($\mathcal{R}_{\theta}$)~\cite{Beaulieu20_2}. We benchmarked this approach by investigating the modulation of photoemission intensity from 2H-WSe$_2$ upon a 60$^{\circ}$ azimuthal crystal rotation. In 2H-TMDCs, due to the broken inversion symmetry within each monolayer and the strong spin-orbit splitting, a 60$^{\circ}$ crystal rotation ($\mathcal{R}_{60^{\circ}}$) is mimicking a time-reversal operation ($\hat{T}$), \textit{i.e.} $\mathcal{R}_{60^{\circ}} \equiv \hat{T}$, which is equivalent to swapping valley-pseudospin indices, \textit{i.e.}, K $\rightarrow$ K$^{\prime}$, and vice-versa. We have shown that since the extrinsic contribution to dichroism was invariant upon $\mathcal{R}_{60^{\circ}}$, the antisymmetric components of the dichroism upon $\mathcal{R}_{60^{\circ}}$ (\textit{i.e.} the differential signal which undergoes a sign change, called Time-Reversal Dichroism in the Photoelectron Angular Distribution - TRDAD) is a background-free intrinsic dichroic signal sensitive to the orbital pseudospin texture in 2H-WSe$_2$~\cite{Beaulieu20_2}. 

In this work, we use the same experimental methodology -- \textit{i.e.} we measure the photoemission intensity modulation upon sample rotation -- to unveil the orbital texture of 1T-TiTe$_2$. We introduce the \emph{intrinsic} linear dichroism in photoelectron angular distributions (\textit{i}LDAD) -- a generalized version of TRDAD, in samples featuring inversion symmetry. Indeed, TRDAD can be seen as a special case of \textit{i}LDAD, when a sample possesses broken inversion symmetry and when the crystal rotation is thus equivalent to a time-reversal operation. The \textit{i}LDAD from 1T-TiTe$_2$, extracted from ARPES measurements for two different crystal orientations, is fully reproduced by first-principle simulations based on the Korringa–Kohn–Rostoker (KKR) method. The excellent agreement of experiment and theory allows us to disentangle different contributions to the ARPES signal and to construct a tight-binding (TB) model capturing the orbital texture. Combining the TB model with qualitative modeling of the photoemission matrix elements, we trace the measured dichroism to the tilted real-space character of the relevant orbitals. The intrinsic dichroism is found to be particularly sensitive to the orbital character, which is beyond standard LDAD based on the anisotropy of the photoemission intensity upon rotating the light polarization direction by 90$^{\circ}$~\cite{Wang12,Sterzi18,Louat19}. Hence, \textit{i}LDAD is a powerful addition to the toolbox of dichroism in ARPES, in the spirit of going beyond band structure mapping and gaining information about energy- and momentum-resolved electronic wavefunction of solids. 

\section{Results}

\subsection{Band structure mapping}

\begin{figure}
\centering\includegraphics[width=0.45\textwidth]{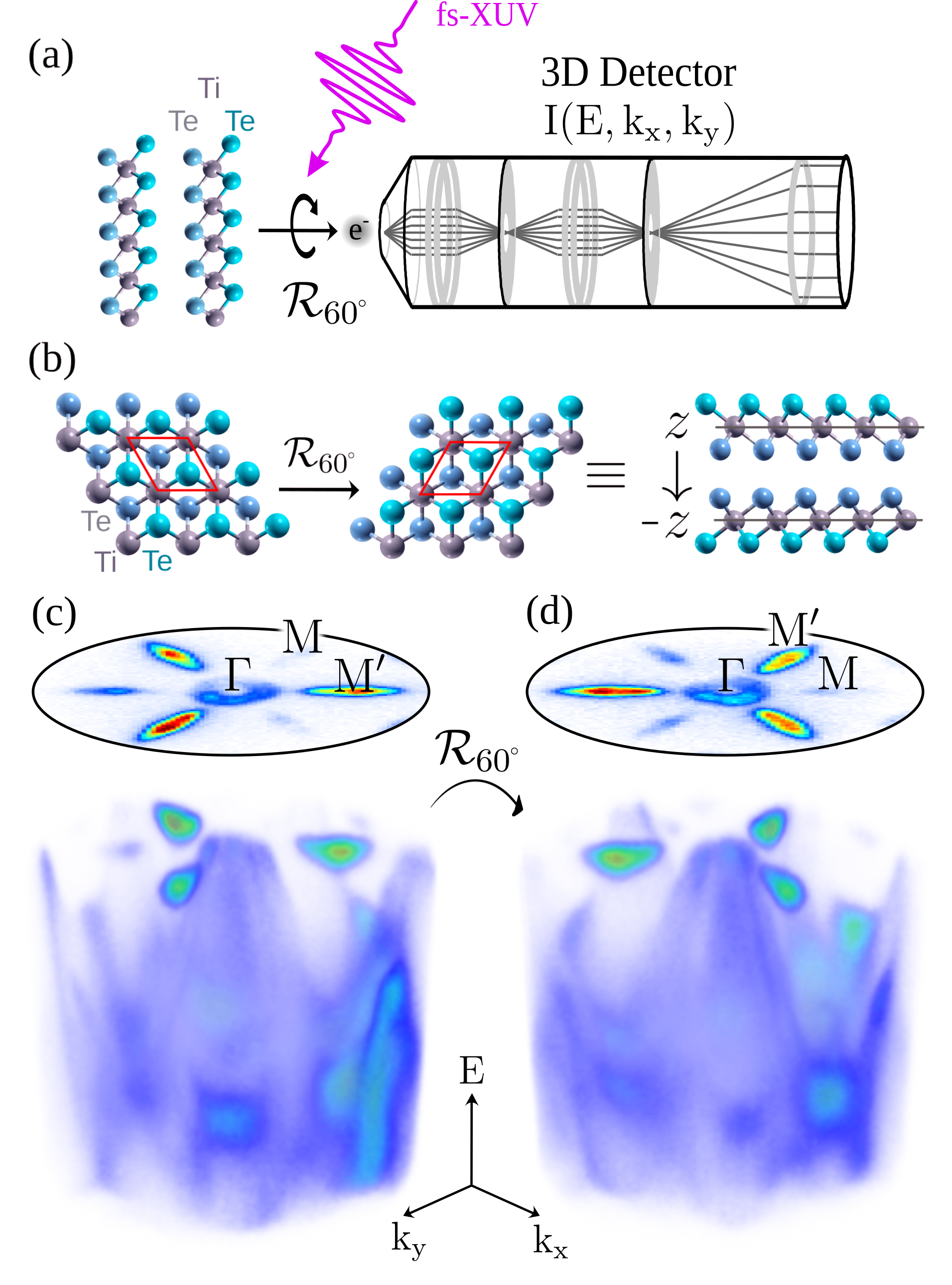}
\caption{\textbf{Scheme of the multidimensional photoemission spectroscopy experiments:}. \textbf{(a)} a $p$-polarized fs-XUV (21.7~eV) pulse is focused onto a freshly cleaved bulk 1T-TiTe$_2$ crystal at an angle of incidence of 65$^{\circ}$ with respect to the surface normal, emitting photoelectrons which are detected by a time-of-flight momentum microscope (METIS1000, SPECS GmbH). The crystal is orientated in the laboratory frame such that the light polarization axis is along the $\Gamma$-M/M' direction. \textbf{(b)} Schematic of the equivalence between a crystal rotation by 60$^{\circ}$ ($\mathcal{R}_{60^{\circ}}$) in the laboratory frame and a mirror operation ($z \rightarrow -z$), in 1T-TiTe$_2$. \textbf{(c)} Three-dimensional photoemission intensity $I(E,k_x,k_y)$, and associated constant energy contour (CEC) at the Fermi energy ($\mathrm{E-E_{F}}$ = 0~eV). \textbf{(d)} Same as \textbf{(c)}, but following a crystal rotation by 60$^{\circ}$ in the laboratory frame.}
\label{fig:intro}
\end{figure}

The experimental apparatus is composed of a table-top femtosecond (fs) XUV source (centered around 21.7~eV, spectral bandwidth 110~meV FWHM)~\cite{Puppin19} coupled to a time-of-flight momentum microscope spectrometer (METIS~1000, SPECS GmbH), see Fig.~\ref{fig:intro}(a). This detector allows measuring all electrons emitted above the surface of the sample and recording a stream of individually detected photoelectrons, resolved in the three intrinsic dimensions of the detector: kinetic energy ($E$) and both parallel momentum components ($k_x$, $k_y$). One of the major advantages of this type of detector is the parallel detection of the 3D photoemission intensity  $I(E,k_x,k_y)$ in a single measurement, without having to rotate the sample in the laboratory frame~\cite{Medjanik17}. More details about the experimental setup can be found elsewhere~\cite{Puppin19,Maklar20} and in the methods. Giving the energy-dependence of the inelastic mean free path of the outgoing electrons~\cite{Seah79,Riley14}, working in the XUV spectral range ensure that most of the detected electrons are coming from the topmost 1T-TiTe$_2$ layer (first atomic trilayer)~\cite{Beaulieu20_2}. More details about this statement, as well as its theoretical validation, will be given later. 

The experimentally measured 3D photoemission intensities, and associated constant energy contours (CECs) at the Fermi energy ($\mathrm{E-E_{F}}$ = 0~eV) are shown in Fig.~\ref{fig:intro}(c)-(d).
Looking at the 3D photoemission intensity (Fig.~\ref{fig:intro}(c)), we recognize the well-known semimetallic nature of 1T-TiTe$_2$, featuring spectrally overlapping Te-5p hole pockets at $\Gamma$ and Ti-3d electron pockets around two inequivalent M points (denoted by M/M')~\cite{Strokov06}. Looking at the CECs in Fig.\ref{fig:intro}(c), the most striking observation is the strong photoemission intensity anisotropy between Ti-3d electron pockets located at M and M'. Indeed, while the photoemission intensity from M' pockets is strong, it is almost completely suppressed for adjacent M pockets. The emergence of this strong linear dichroism in photoemission \textit{a priori} can originate from extrinsic (\textit{e.g.} experimental geometry) and/or intrinsic (\textit{i.e.} orbital/Bloch wavefunction symmetries) effects. 

Motivated by the recently introduced approach for disentangling extrinsic and intrinsic contributions to dichroism in ARPES (see ref.~\cite{Beaulieu20_2} and section~\ref{intro}), we also performed band structure mapping of 1T-TiTe$_2$ for a second crystal orientation rotated by 60$^{\circ}$. In inversion symmetric 1T-TiTe$_2$, this crystal rotation ($\mathcal{R}_{60^{\circ}}$) is strictly equivalent to a mirror-image symmetry of each layer (see Fig.~\ref{fig:intro}(b)), \textit{i.e.} $\mathcal{R}_{60^{\circ}} \equiv z\rightarrow-z$. Although the effect of a 60$^{\circ}$ crystal rotation in 1T-TMDC ($\mathcal{R}_{60^{\circ}} \equiv z\rightarrow-z$) is different than in 2H-TMDC ($\mathcal{R}_{60^{\circ}} \equiv \hat{T}$), in both cases it corresponds to a well-defined symmetry operation on the electronic wavefunction of the crystal. Linking a given symmetry operation on the electronic wavefunction to the modulation of the photoemission intensity, complemented by proper theoretical analysis, is a powerful tool to investigate the energy- and momentum-dependent orbital texture, a quantity of fundamental importance but hardly accessible with other techniques. 

\subsection{Intrinsic linear dichroism in photoelectron angular distributions}

\begin{figure}
\centering\includegraphics[width=0.45\textwidth]{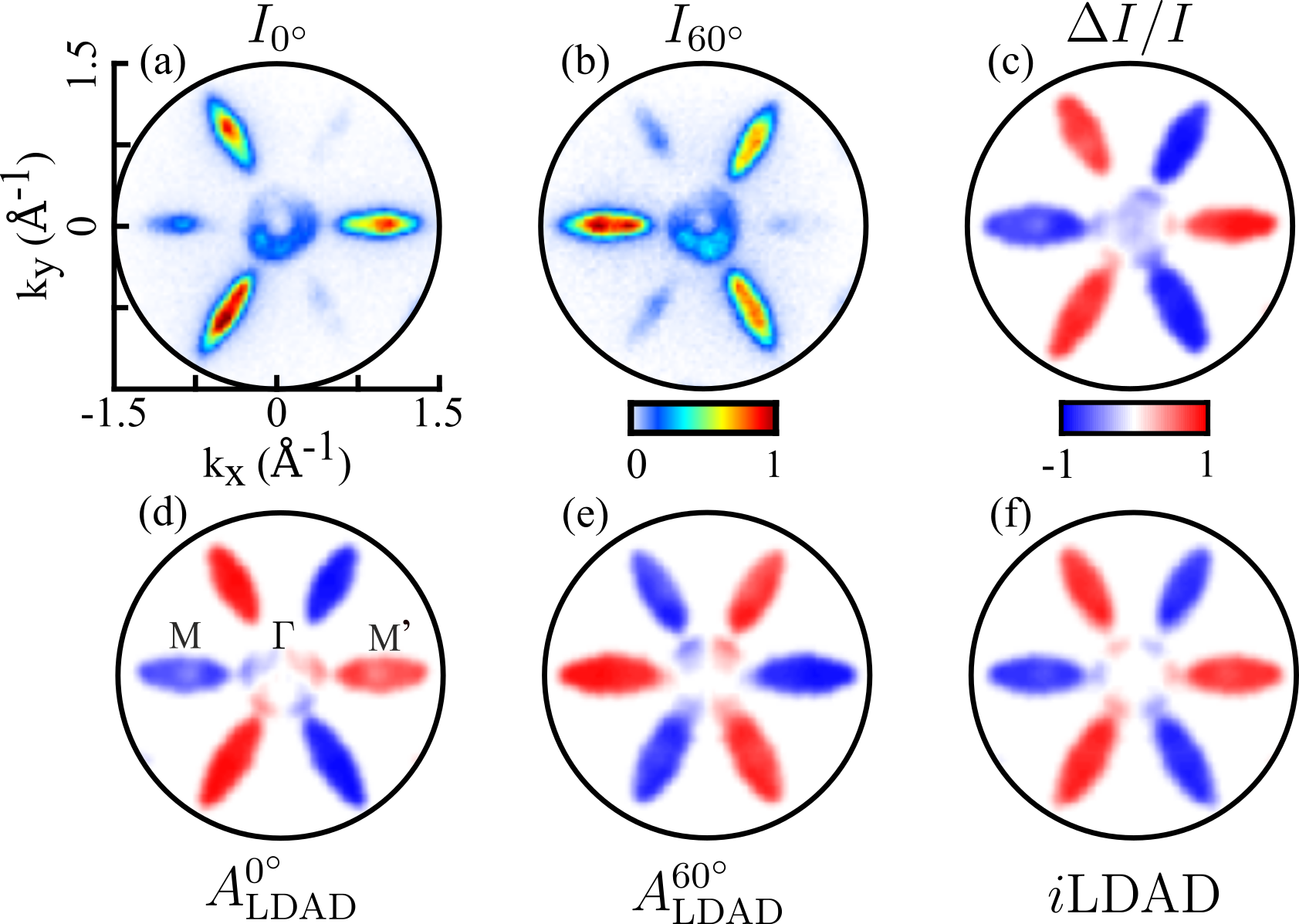}
\caption{\textbf{Extraction of the \textit{intrisic} Linear Dichroism in Photoelectron Angular Distributions (\textit{i}LDAD)}: \textbf{(a)-(b)} $I_{0^{\circ}}$ and $I_{60^{\circ}}$, the CECs at the Fermi energy, measured for two crystal orientations rotated by 60$^{\circ}$ with respect to each others. \textbf{(c)} $\Delta I/I$, the raw normalized difference, \textit{i.e.} ($I_{0^{\circ}}$ - $I_{60^{\circ}}$)/($I_{0^{\circ}}$ + $I_{60^{\circ}}$) between CECs shown in (a) and (b). \textbf{(d)-(e)} $A_{LDAD}^{0^{\circ}}$ and $A_{LDAD}^{60^{\circ}}$, the “left-right asymmetries”, reflect the photoemission intensity asymmetry between the $k_x\!<$0 and $k_x\!>$0 hemispheres, for two crystal orientations, respectively. \textbf{(f)} \textit{i}LDAD represents the component of $A_{LDAD}^{0^{\circ}/60^{\circ}}$ which is antisymmetric upon 60$^{\circ}$ azimuthal rotation of the crystal.}
\label{fig:ldad_exp}
\end{figure}

In Fig.~\ref{fig:ldad_exp}, we show the procedure to extract the antisymmetric components of the dichroism (referred to as \textit{intrinsic} Linear Dichroism in Photoelectron Angular Distributions - \textit{i}LDAD) from photoemission data obtained from two crystal orientations rotated by 60$^{\circ}$ with respect to each other. In Fig.~\ref{fig:ldad_exp}(a), one can see the CEC at the Fermi energy featuring the triangular doughnut-shaped Te-5p hole pockets around $\Gamma$, which shows subtle but non-vanishing angular anisotropies. More remarkably, the strongly anisotropic photoemission from Ti-3d electron pockets located around M and M' is dominating the signal. As already discussed in the context of Fig.~\ref{fig:intro}, while the photoemission intensity is strong for the M' pockets, it is almost vanishing for the adjacent M pockets. Upon a 60$^{\circ}$ azimuthal rotation of the crystal (see Fig.~\ref{fig:ldad_exp}(b)), this asymmetric photoemission yield from M and M' pockets remained mostly unchanged, leading to a 60$^{\circ}$ rotation of the momentum-resolved photoemission patterns. This is a strong indication that this asymmetric photoemission yield is of intrinsic origin and should encode information about the symmetry properties of the orbitals. 

While Fig.~\ref{fig:ldad_exp}(b) shows the raw normalized photoemission intensity difference between the two crystal orientations, using our multidimensional detection scheme with the $p$-polarized fs-XUV pulses impinging in the $x$-$z$ plane (along $\Gamma$-M/M'), the normalized intensity differences between the forward $I_{\alpha}(E,k_x,k_y)$ and backward $I_{\alpha}(E,-k_x,k_y)$ hemisphere, \textit{i.e.}~the linear dichroism asymmetry in the photoelectron angular distribution $A_{LDAD}^{\alpha}(E,k_x,k_y)$, can be extracted (see Eq.~\ref{Eq:A_LDAD}), without the need to rearrange the sample geometry nor the light-polarization state~\cite{Chernov15,Tusche16}. 
    
\begin{equation}
    A_{LDAD}^{\alpha} =
    \frac{I_{\alpha}(E,k_x,k_y) - I_{\alpha}(E,-k_x,k_y)}{I_{\alpha}(E,k_x,k_y) + I_{\alpha}(E,-k_x,k_y)}
\label{Eq:A_LDAD}
\end{equation}

Looking at $A_{LDAD}^{0^{\circ}}$ (Fig.~\ref{fig:ldad_exp}(d)), we notice very strong dichroism (up to $\pm$97$\%$) around M/M' pockets with an opposite sign for adjacent valleys, and weaker dichroism (up to $\pm$35$\%$) with the same threefold symmetry and sign texture for the Te-5p hole pockets around $\Gamma$. The dichroic signal is not perfectly up/down ($k_y>0$/$k_y<0$) symmetric, suggesting small experimental imperfection in the crystal orientation, XUV beam alignment, or photoelectron imaging, since the up/down symmetry should not be broken within this experimental configuration (\textit{i.e.} $p$- polarization along $\Gamma$-M/M'). Moreover, since we are dealing with normalized signals (see Eq.~\ref{Eq:A_LDAD}), we only show the dichroism for momentum regions where the right-left ($k_x>0$/$k_x<0$) symmetrized signal is $>0.1I_{max}$, where $I_{max}$ is the maximum photoemission intensity for this CEC. This thresholding procedure ensures that we do not over-interpret the potentially strong dichroism coming from momentum regions of vanishing photoemission intensity. Fig.~\ref{fig:ldad_exp}(e) shows $A_{LDAD}^{60^{\circ}}$, \textit{i.e.} the dichroism extracted after rotating the crystal orientation by 60$^{\circ}$. The measured $A_{LDAD}^{60^{\circ}}$ dichroic signal is almost perfectly antisymmetric with respect to $A_{LDAD}^{0^{\circ}}$, again indicating its intrinsic origin since an extrinsic (experimental geometry) origin would have led to a dichroic signal invariant upon a $\mathcal{R}_{60^{\circ}}$ operation. 

In order to quantitatively isolate the antisymmetric part of the $A_{LDAD}^{\alpha}$ dichroism upon crystal rotation ($\mathcal{R}_{60^{\circ}}$), \textit{i.e.} to remove any spurious contributions from experimental geometry, we introduce the \textit{i}LDAD: 

\begin{equation}
    \textit{i}LDAD = \frac{A_{LDAD}^{\alpha} -  A_{LDAD}^{\alpha'}}{2}
    \label{Eq:A_LDAD_T}
\end{equation}

where the crystal rotation is hereby an angle $\alpha$-$\alpha'$ = 60$^{\circ}$. The resulting \textit{i}LDAD is shown in Fig.~\ref{fig:ldad_exp}(f). The strong similarity between $A_{LDAD}^{0^{\circ}}$ and \textit{i}LDAD is another observation revealing that the dichroism in 1T-TiTe$_2$ is almost exclusively originating from intrinsic origins (orbital symmetry/orientation). We will investigate the microscopic origin of the emergence of this strong intrinsic dichroism and how it can be used to reveal key information about the orbital texture of this layered semimetallic TMDC. 

\subsection{Comparison with DFT band structure and one-step model of photoemission}

To extract quantitative information about orbital texture from such differential (\textit{i}LDAD) ARPES measurements, a systematic comparison with quantitative theories is highly beneficial. Here, we use inputs from two different theoretical approaches in a synergistic manner. First, we performed one-step model photoemission calculations based on fully relativistic density functional theory (DFT). The one-step model of photoemission is implemented in the fully relativistic KKR method \cite{ebert_calculating_2011,braun_correlation_2018}. The calculated photoemission signal includes all matrix element effects such as experimental geometry, photon energy, polarization state, and final state effects (for more details, see methods). Second, we constructed a first-principle TB model by computing projective Wannier functions (see methods). The TB model leads to an excellent description of the electronic structure of 1T-TiTe$_2$ and provides an intuitive picture of the link between orbital degrees of freedom and dichroism in photoemission.

\begin{figure}
\centering\includegraphics[width=0.45\textwidth]{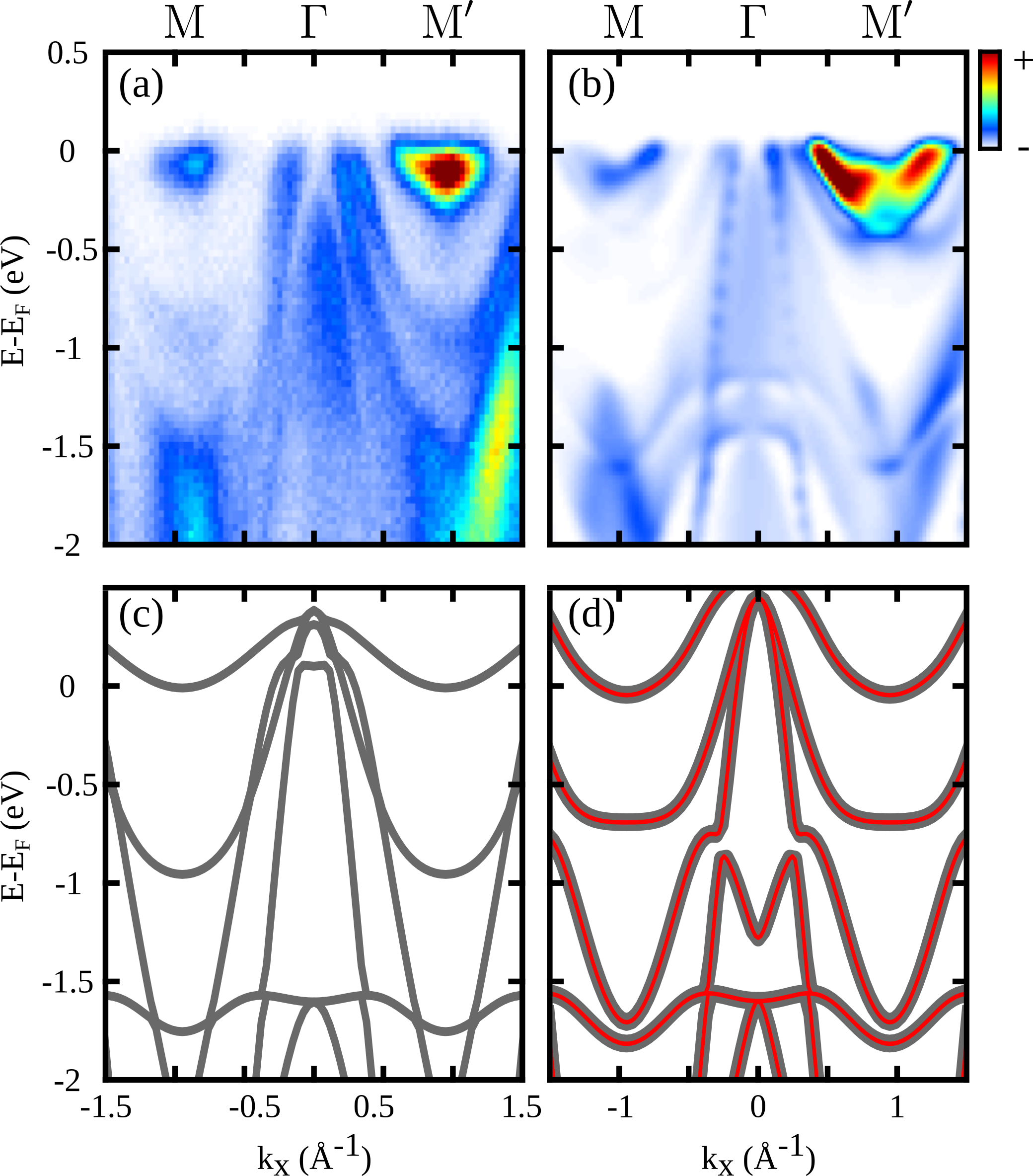}
\caption{\textbf{Band structure of 1T-TiTe$_2$ along the M-$\mathrm{\Gamma}$-M' cut}. \textbf{(a)} Experimentally measured photoemission intensity using the same experimental geometry as described in Fig.~\ref{fig:intro}. \textbf{(b)} Photoemission intensity along the same cut obtained using the one-step model (KKR) method. 
The electronic structure has obtained from the local density approximation (LDA).
\textbf{(c)} Band structure of bulk TiTe$_2$ obtained from nonrelativistic DFT calculations. We used the Perdew–Burke-Ernzerhof (PBE) functional.
\textbf{(d)} Band structure of a free-standing monolayer from DFT calculations (grey lines). The thin red lines represent the bands obtained from the \textit{ab initio} TB model, which has been constructed from projective Wannier functions. }
\label{fig:bands}
\end{figure}

An important step, before calculating \textit{i}LDAD, is to ensure that both theoretical approaches can accurately capture the electronic structure of 1T-TiTe$_2$. The experimentally measured photoemission intensity, the simulated KKR photoemission intensity, and the calculated DFT band structure along the M-$\mathrm{\Gamma}$-M' cut are shown in Fig.~\ref{fig:bands}. The experimental measured photoemission intensity are shown in Fig.~\ref{fig:bands}(a), from which one can clearly recognize the well-known semimetallic nature of 1T-TiTe$_2$, featuring Te-$5p$ hole pockets at $\Gamma$ and Ti-$3d$ electron pockets around M/M' points.  The experimentally obtained band structure is well captured by the calculated DFT band structure. Within the local density approximation (LDA) -- including spin-orbit coupling (SOC) -- used for the KKR method (Fig.~\ref{fig:bands}(b)), the measured band structure near the Fermi level is well reproduced, both at the $\Gamma$ point and at the M/M' pockets. At lower energies (E-E$_\mathrm{F}\sim -0.9$~eV) the measured spectrum shows bands that are absent in the calculated spectrum. We also performed DFT calculations using the Perdew–Burke-Ernzerhof (PBE) functional (Fig.~\ref{fig:bands}(c)), which captures these bands. The size of the pockets at M/M' is slightly smaller than in the experiments and within the LDA, which is consistent with calculations including exchange effects~\cite{zhou_theory_2020}. We have also calculated the band structure of a free-standing monolayer (Fig.~\ref{fig:bands}(d)) for comparison. Albeit the band structure is generally modified with respect to the bulk (especially near $\Gamma$), the electron pockets at M/M' are still present. Constructing a TB model from projective Wannier functions (red lines in Fig.~\ref{fig:bands}(d)) provides an excellent fit of PBE band structure. The TB description provides us with a simple model for understanding the measured photoemission features at M/M', which is discussed below.
 
By looking more carefully at Fig.~\ref{fig:bands}(a), and as seen in Fermi level CECs (Fig.~\ref{fig:intro}(c)-(d)), one can notice strong photoemission anisotropies between positive and negative $k_x$ (corresponding to $\mathrm{\Gamma}$-M' and $\mathrm{\Gamma}$-M direction, respectively). These photoemission anisotropies, originating from transition dipole matrix element effects, are well-captured by the calculated KKR photoemission intensity, shown in Fig.~\ref{fig:bands}(b). As mentioned earlier, the most striking asymmetry in the photoemission signal is coming from  Ti-3d electron pockets around M/M' points, near the Fermi level. Moreover, because the role of orbital texture near the Fermi level is of capital importance in determining materials' optical and transport properties, we will focus the subsequent analysis on the photoemission anisotropies near the Fermi energy. 

\subsection{Calculated \textit{i}LDAD and its photon-energy dependence using KKR\label{subsec:kkrcalc}}

\begin{figure}
\centering\includegraphics[width=0.45\textwidth]{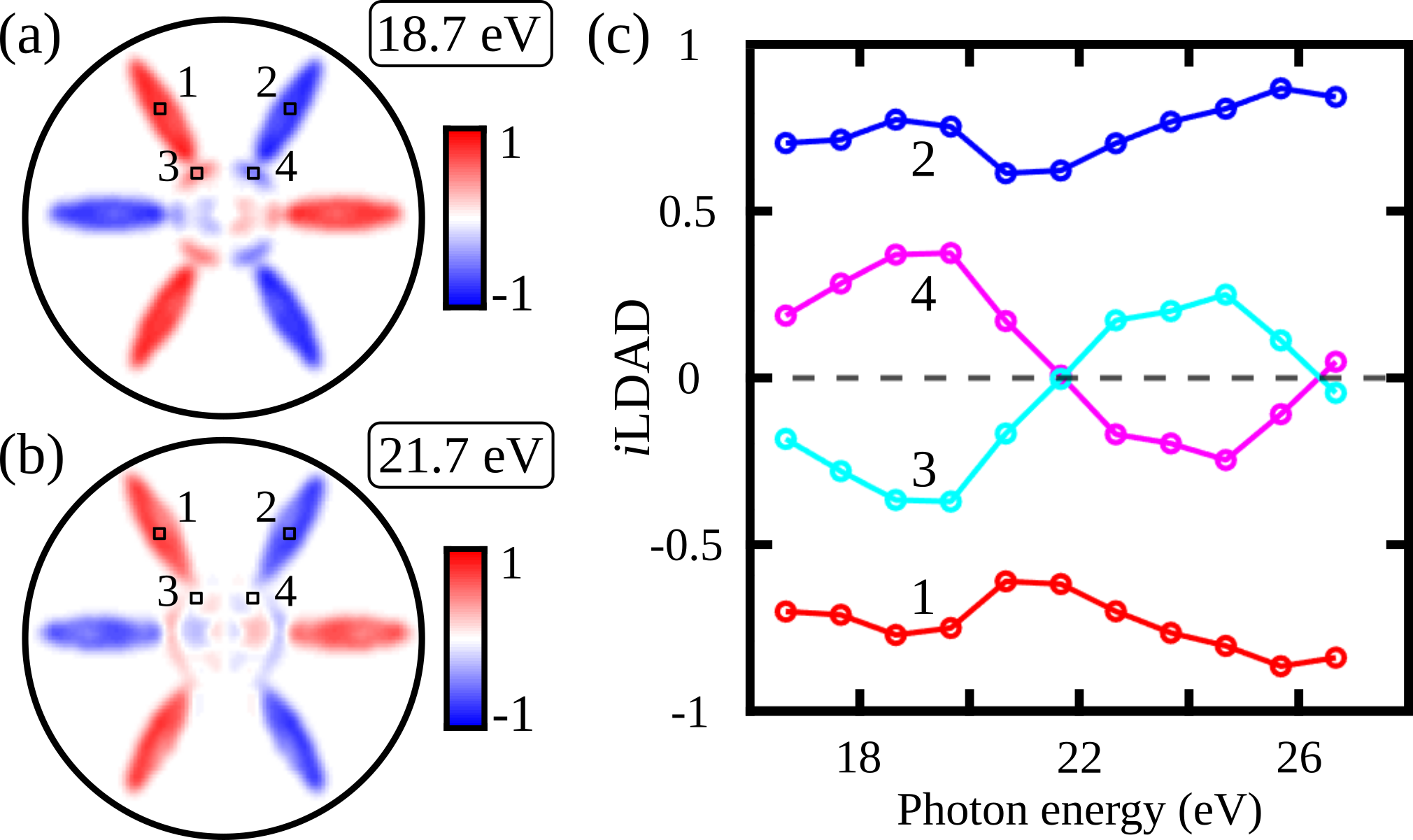}
\caption{\textbf{Photon energy dependence of \textit{i}LDAD calculated using the KKR method.} \textbf{(a)-(b)} \textit{i}LDAD calculated using the KKR and a photon energy of 18.7 eV. The regions of interest label 1-4 are selected to investigate the photon energy dependence of \textit{i}LDAD \textbf{(b)} \textit{i}LDAD as a function of photon energy, in the selected region of interest shown in \textbf{(a)}}
\label{fig:photon_dep}
\end{figure}

In Fig.~\ref{fig:photon_dep} (a) and (b) we compare the \textit{i}LDAD calculated using the one-step model (KKR) for two different photon energies (18.7~eV and 21.7~eV, respectively). For both photon energies, the calculated \textit{i}LDAD at M/M' pockets is very similar to the experimentally measured \textit{i}LDAD. However, surprisingly, using 18.7~eV instead of the experimentally used 21.7~eV photon energy captures better the \textit{i}LDAD features around $\Gamma$. This can be interpreted as a limitation of DFT: it is well known that the calculated time-reversed low-energy electron diffraction (TRLEED) final state can be shifted in energy with respect to the experimental value (for more details see \ref{sec:methods}). However, the dispersion and the orbital character of the time-reversed LEED state are typically well captured. Moreover, it was pointed out by Strocov et al.~\cite{Strokov06} that for 1T-TiTe$_2$ the photoemission final state in the UV photon energy regime features dramatic final state effects such as multiband composition and non-free electron like dispersions around the center of the Brillouin zone. In addition, it was shown that the states around $\Gamma$ at the Fermi level show three-dimensional character due to the interactions between 1T-TiTe$_2$ monolayers (atomic trilayers)~\cite{Strokov06}. 

To investigate the role of the final states, we performed a detailed theoretical photon-energy-dependent \textit{i}LDAD scan ($k_z$-scan). In Fig. \ref{fig:photon_dep} (c), we show the photon energy dependence of the \textit{i}LDAD in the selected regions of interest (marked as 1-4). While the \textit{i}LDAD of Ti-3d derived electron pockets around M/M' points (features 1 and 2) are extremely stable against variation of the photon energy, the hole pockets around $\Gamma$ point (features 3 and 4) exhibits strong modulations including even sign reversal upon modification of the photon energy. We attribute this to the strong 3D character of Te-5p states. A similar trend was observed for the soft-X-ray photoemission, where the final state becomes truly free-electron-like (see SI), ruling out the predominance of final-state effects in the emergence of \textit{i}LDAD.

\subsection{The role of orbital texture in photoemission: tight-binding analysis \label{subsec:tbarpes}}

While the \textit{i}LDAD exhibits a pronounced photon energy dependence near the $\Gamma$ point,
the robustness of the \textit{i}LDAD near the M/M' pockets gives us confidence that we can link its origin to microscopic properties of the electronic structure of 1T-TiTe$_2$. To find an intuitive link between the orbital texture and this intrinsic dichroic signal, we employ an \textit{ab initio} TB model. Within this model the Bloch wavefunction is represented by
\begin{align}
\label{eq:blochwf}
  \psi_{\vec{k}n}(\vec{r}) = \frac{1}{N} \sum_{\vec{R}} e^{i \vec{k}\cdot\vec{R}}\sum_j C_{j n}(\vec{k}) \phi_j(\vec{r} - \vec{R})  ,
\end{align}
where $\phi_j(\vec{r})$ denote the Wannier functions localized at particular sites $\vec{R}$. The coefficients $C_{jn}(\vec{k})$ are the coefficients connection orbital (index $j$) and band (index $n$) space; they are the eigenvectors of the TB Hamiltonian $H^\mathrm{TB}_{jj^\prime}(\vec{k})$. The Wannier representation~\eqref{eq:blochwf} provides a direct of real and reciprocal space. For each momentum $\vec{k}$ the Bloch wavefunction is a coherent superposition of the Wannier functions $\phi_j(\vec{r})$, whose orbital symmetry is directly reflected in the ARPES signal.

\begin{figure*}
\includegraphics[width=1\textwidth]{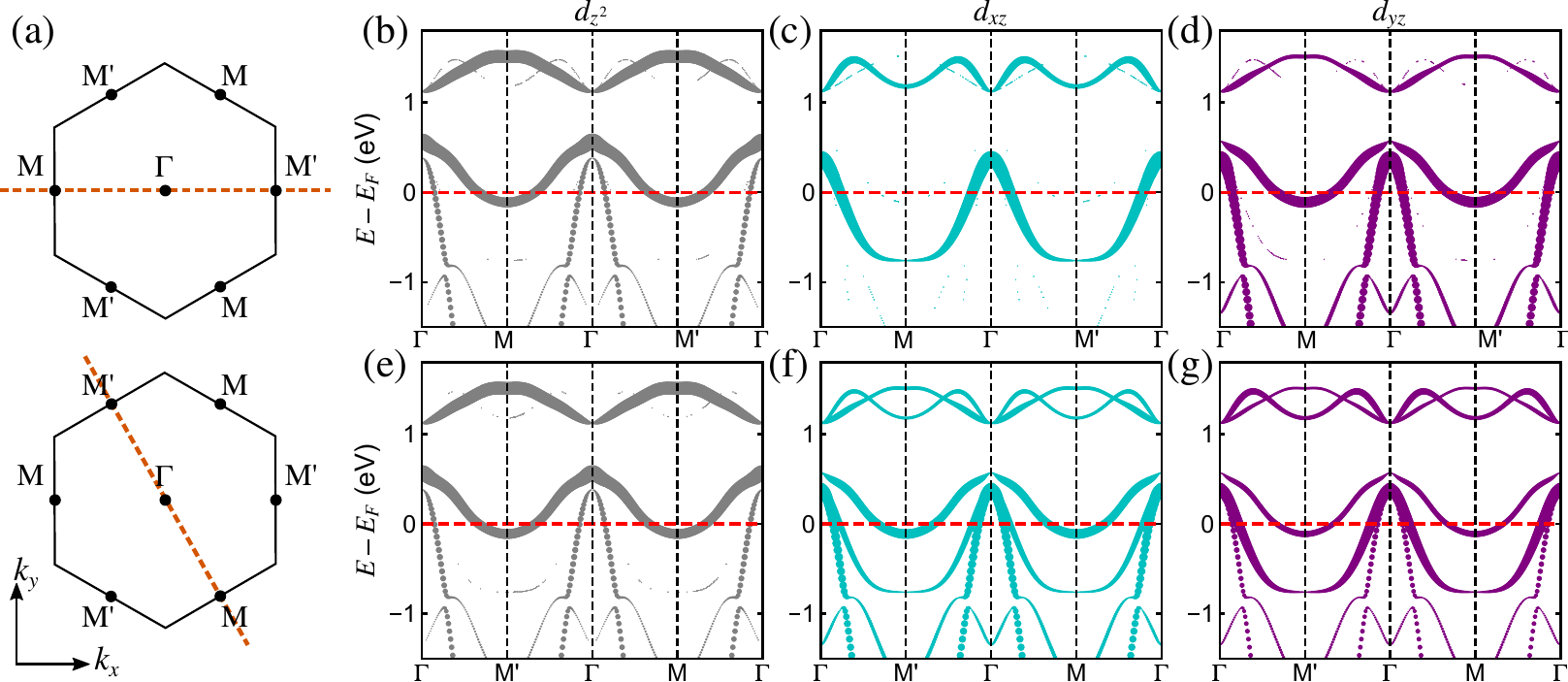}
\caption{\textbf{Orbital-projected band structure of monolayer 1T-TiTe$_2$.} \textbf{(a)} First surface Brillouin zone (black) and momentum cut (dashed lines). \textbf{(b)-(d)} 
Fat-band representation of the band structure projected onto the $d_{z^2}$, $d_{xz}$ and $d_{yz}$ orbitals. The line thickness represents the weight of the respective orbital. \textbf{(e)-(g)} Fat-band representation of the orbital weight for the lower cut in (a).
\label{fig:orbital_bands}}
\end{figure*}

From a layer-resolved KKR calculation, we found that the characteristic intensity variation can be captured by including the top layer TiTe$_2$ only (see SI). Therefore, we constructed the TB model for a free-standing monolayer, including five Ti-$d$ orbitals three $p$ orbitals for each of the two Te chalcogen atoms in Eq.~\eqref{eq:blochwf}. The model provides a systematic way to examine the orbital character of the bands; in particular, we find that the contribution of the Te-$p$ orbitals is negligible near E$_\mathrm{F}$ around M/M' pockets (see SI). Fig.~\ref{fig:orbital_bands} shows the corresponding orbital-projected band structure for the dominant Ti-$d$ orbitals along the M-$\Gamma$-M' (upper row) and a M'-$\Gamma$-M cut rotated by 60$^\circ$ (lower row in Fig.~\ref{fig:orbital_bands}), respectively. Focusing on the electron pockets crossing the Fermi level near M/M', it is clear that the out-of-plane Ti-$d_{z^2}$ orbitals (Fig.~\ref{fig:orbital_bands}(b,e)) contribute equally to the density of state along both paths. Inspecting the Ti-$d_{xz}$ and Ti-$d_{yz}$ orbital-projected band structure (Fig.~\ref{fig:orbital_bands}~(c--d) and (f--g) , respectively), the situation is drastically different. Along the path parallel to $k_x$ the weight of the Ti-$d_{xz}$ orbital is negligible at the M/M' pockets, while the Ti-$d_{yz}$ orbital contributes significantly. Along the rotated path the band at the M/M' pocket is comprised of the Ti-$d_{z^2}$ and the Ti-$d_{xz}$ orbital with almost equal weight, with a smaller contribution from the Ti-$d_{yz}$ orbital. It is important to note that while the band structure possesses six-fold rotational symmetry in momentum space, the orbital texture exhibits a three-fold symmetry corresponding to the lattice symmetry. 

\begin{figure}
\centering\includegraphics[width=0.5\textwidth]{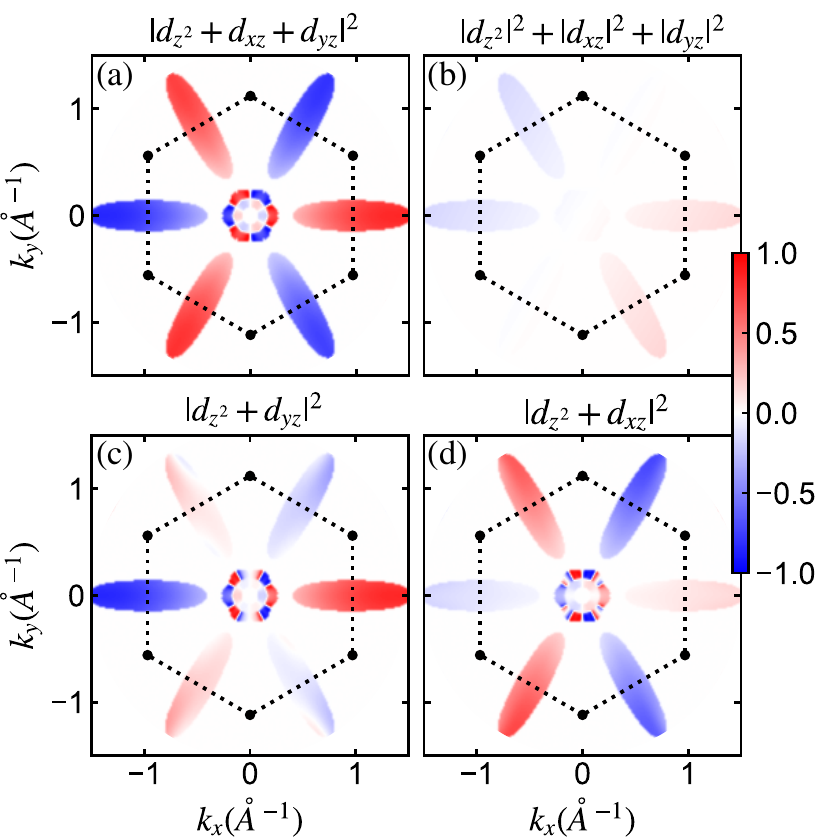}
\caption{\textbf{Role of orbital mixing in photoemission.} \textbf{(a)} \textit{i}LDAD calculated using the TB+PW model where the initial state is composed of a coherent superposition of $d_{z^2}$, $d_{xz}$ and $d_{yz}$ orbitals. \textbf{(b)} Same as \textbf{(a)}, for the initial state being an incoherent sum the signal including the $d_{z^2}$, $d_{xz}$ or $d_{yz}$ orbitals.  \textbf{(c)} Same as \textbf{(a)} but including only $d_{z^2}$ and $d_{yz}$ orbitals. \textbf{(d)} Same as \textbf{(a)} for the initial state being a coherent combination of $d_{z^2}$, and $d_{xz}$ orbitals. The binding energy in all plots is fixed at $E_B=-0.05$~eV; the photon energy is $\hbar \omega = 21.7$~eV as in the experiments.}
\label{fig:orbital_ldad}
\end{figure}

This momentum-dependent orbital content of the band structure underlines the important role of the orbital degree of freedom in photoemission from 1T-TiTe$_2$. For exploring the manifestations in the ARPES signal, we combined the TB model with modeling the Wannier functions and the plane-wave (PW) approximation to the final photoelectron states (see methods).  We calculated the photoemission spectrum within the resulting TB+PW model for both sample orientations (see methods) and extracted the \textit{i}LDAD. Including specific orbitals only then allows for isolating the individual channels giving rise to the measured \textit{i}LDAD.

\begin{figure*}
\includegraphics[width=1\textwidth]{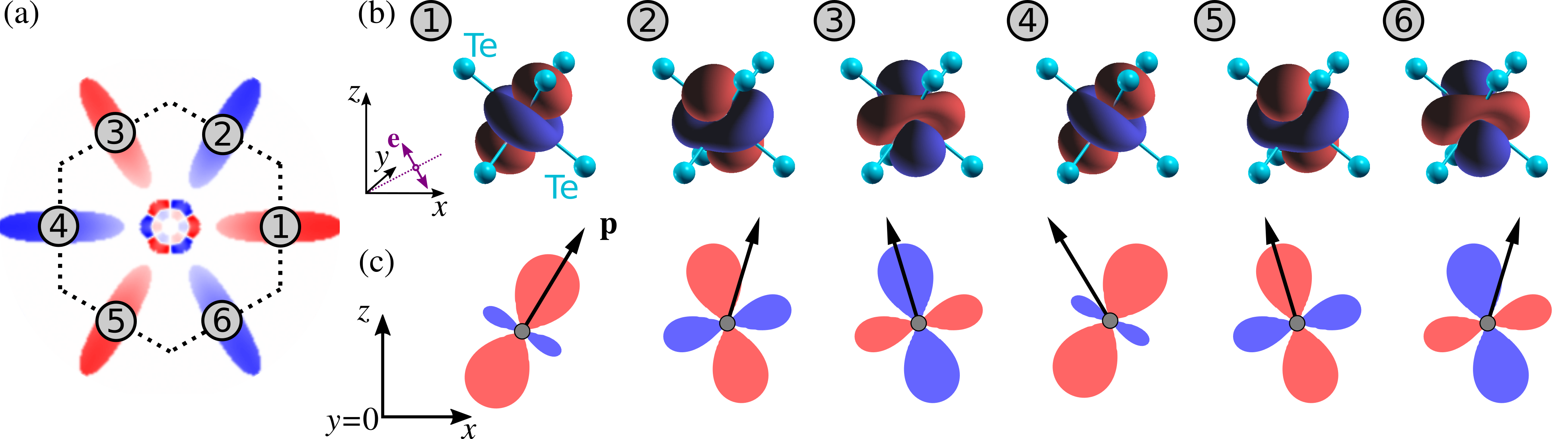}
\caption{\textbf{Physical origin \textit{i}LDAD: Momentum-dependent orbital orientation} \textbf{(a)} Sketch of the calculated (TB+PW) \textit{i}LDAD with labeled M/M' pockets (1-6). \textbf{(b)} Sketch of the hybrid orbital $\widetilde{\psi}_m(\vec{r})$, for each M/M' points shown in \textbf{(a)}. The orbital is represented by the constant-value contour $|\widetilde{\psi}_m(\vec{r})| = 0.02$, while the red (blue) color shading indicates $\mathrm{Re}[\widetilde{\psi}_m(\vec{r})] > 0$ ($\mathrm{Re}[\widetilde{\psi}_m(\vec{r})] < 0$).  The coordinate systems corresponds to the laboratory frame. \textbf{(c)} projection of the hybrid orbitals onto the $x$--$z$ plane. The photoelectron wave vector $\vec{p}$ is indicated by the black arrow.}
\label{fig:origin} 
\end{figure*}

First, in Fig.~\ref{fig:orbital_ldad}(a), we show the \textit{i}LDAD obtained from the TB+PW model including the Ti-$d_{z^2}$, $d_{xz}$ and $d_{yz}$ orbitals. The calculated \textit{i}LDAD is in excellent qualitative agreement with the measured signal near the M/M' pockets, which confirms that these three orbitals are the main ingredients of the \textit{i}LDAD signal. 
Moreover, the coherent superposition of these orbitals is important: when incoherently summing $i\mathrm{LDAD} = i\mathrm{LDAD} _{z^2} + i\mathrm{LDAD}_{xz} + i\mathrm{LDAD}_{yz}$ 
, \textit{i}LDAD is extremely weak and does not exhibit the expected symmetry (as experimentally measured), as shown in Fig. \ref{fig:orbital_ldad}(b). Including only the $d_{z^2}$ and $d_{yz}$ orbitals (Fig.~\ref{fig:orbital_ldad}(c)) captures the full signal along the $k_x$ axis, which is consistent with the corresponding orbital weight (see Fig.~\ref{fig:orbital_bands}). In contrast, including only the $d_{z^2}$ and $d_{xz}$ orbitals (Fig.~\ref{fig:orbital_ldad}(d)) accounts for most of the \textit{i}LDAD for the upper and lower two pockets, respectively. Note that the Ti-$d$ orbitals are defined in the laboratory coordinate system here, which is the natural choice when computing ARPES signals. Earlier experiments~\cite{smith_angular-resolved_1975,traum_angular_1974} attempted to connect the photoemission signal to the symmetry-adapted crystal-field orbitals. The crystal axes defined by the Ti-Te bonds are not strictly orthogonal for the presented experimental geometry, which complicates this picture. In the SI, we present a similar orbital analysis as above in the crystal-field basis for a complementary picture. 

The TB+PW model allows for an even more intuitive picture for the characteristic \textit{i}LDAD by inspecting the real-space dependence of the wavefunction. As explained in the methods section, the photoemission matrix elements can be expressed in terms of the Wannier functions (for the Ti-$d$ orbitals) and eigenvectors $C_{jn}(\vec{k})$. Directly at the six M/M' points on the edge of the Brillouin zone (labeled by $\vec{k}_m, m=1,\dots, 6$; see Fig.~\ref{fig:origin}(a)), the matrix element within the TB+PW model reduces to
\begin{align}
    \label{eq:mel_hyb}
    M(\vec{k}_m,E) = \int d \vec{r}\, e^{-i \vec{p} \cdot \vec{r}} \vec{e}\cdot \vec{r} 
    \widetilde{\psi}_m(\vec{r}) \ ,
\end{align}
where $\vec{e}$ denotes the polarization vector, while $\vec{p}=(k_{m,x},k_{m,y},p_\perp)$ is the three-dimensional momentum of the photoelectron. The \emph{hybridized} initial state $\widetilde{\psi}_m(\vec{r})$ is given by the superposition
\begin{align}
    \label{eq:hybridwf}
    \widetilde{\psi}_m(\vec{r}) = \sum_{j} C_{jn}(\vec{k}_m) \phi_j(\vec{r}) \ ,
\end{align}
where the band index $n$ corresponds to the band near $E_F$.  
The magnitude of the matrix element~\eqref{eq:mel_hyb} is thus determined by the hybrid orbital~\eqref{eq:hybridwf}, in particular its orientation with respect to the photoelectrom momentum $\vec{p}$. The hydrid orbital~\eqref{eq:hybridwf} corresponds to taking the $\vec{R}=0$ contribution in Eq.~\eqref{eq:blochwf}. Including only the well-localized Ti-$d$ orbitals, we find $\widetilde{\psi}_m(\vec{r})\approx N\psi_{\vec{k_m} n}(\vec{r})$ (up to a trivial phase) in the unit cell; the hybrid orbital~\eqref{eq:hybridwf} is thus a good representation of the Bloch wavefunction.

In Fig.~\ref{fig:origin}(b) we plot $\widetilde{\psi}_m(\vec{r})$ for the $\alpha=0^{\circ}$ sample orientation. Inspecting Fig.~\ref{fig:origin}(b) lead us to the following observation: the electronic wavefunction can be qualitatively described by tilted $d_{z^2}$ orbital, where the tilt angle is determined by the coefficients $C_{jn}(\vec{k}_m)$ and is thus momentum-dependent. This rotated $d_{z^2}$ is mainly composed of the $d_{z^2}$, $d_{xz}$ and $d_{yz}$ orbitals in the laboratory frame. By projecting the hybrid wavefunction at each M/M' point onto the $x$-$z$ plane and comparing their relative orientation to the outgoing electron momentum $\vec{p}$, the origin of the strong \textit{i}LDAD becomes clear. For $m=1,3,5$ the lobes of $\widetilde{\psi}_m(\vec{r})$ align with $\vec{p}$, leading to enhanced photoemission intensity, while for $m=2,4,6$ the photoelectron momentum is roughly pointing along the nodal direction of the hybrid orbitals. Rotating by 60$^\circ$ or equivalently transforming $z\rightarrow -z$ inverts this behavior, as the alignment of $\vec{p}$ with the orientation of the lobes (nodal direction) transforms into alignment with the nodal direction (orientation of the lobes). 

The \textit{i}LDAD is thus sensitive to the orbital orientation. In contrast to regular linear dichroism, the sample rotation $\mathcal{R}_{60^\circ}$ changes the relative orientation of photoelectron momentum and orbitals while keeping the light polarization fixed. Revisiting Eq.~\eqref{eq:mel_hyb}, the photoemission matrix element (within the PW approximation) in essence is a Fourier transform of the hybrid orbital. From the momentum-space dependece of \textit{i}LDAD, we gain insights into the real-space orientation of the orbitals. This information is complementary to varying the light polarization, where the matrix element is determined by the selection rules. In particular, orbital tilting as in 1T-TiTe$_2$ is hard to discern directly from standard linear dichroism -- when comparing the photoemission intensity modulation upon swapping between $s-$ and $p-$ polarized light. 

\section{Discussion}

We have extended the newly introduced differential measurement methodology in multidimensional photoemission spectroscopy to investigate the role of orbital texture in photoemission from a semimetallic layered transition metal dichalcogenide (1T-TiTe$_2$). Indeed, we introduced the \emph{intrinsic} linear dichroism in photoelectron angular distributions (\textit{i}LDAD) -- a generalized version of TRDAD, in samples featuring inversion symmetry. By tracking the photoemission intensity modulation upon specific crystal rotation in the laboratory frame, we experimentally extracted the intrinsic linear dichroism in the photoelectron angular distribution. Comparing our results with two different theoretical calculations based on a quantitative one-step model of photoemission and an intuitive tight-binding model, we provide an intuitive interpretation for the link between orbital texture and the pronounced photoemission intensity anisotropies from momentum regions with equivalent energy eigenvalues but with nonequivalent electronic wavefunctions. Our joint experimental and theoretical work demonstrated the direct relation between momentum-dependent hybridized orbital orientations and photoemission intensities, unambiguously ruling out final state effects as the primary origin of the observed  dichroism~\cite{smith_angular-resolved_1975,traum_angular_1974}. 

The methodology that we used is general and can be exploited to investigate a plethora of phenomena governed by orbital physics. Indeed, one expects very similar types of orbital texture to be common in governing properties of similar TMDC exhibiting multi-component charge order (\textit{e.g.} 1T-TiSe$_2$~\cite{Salvo76} and 1T-VSe$_2$~\cite{Eaglesham86}), so that our methodology can be used to investigate the link between orbital texture of the emergence of charge density waves~\cite{peng_observation_2021,ritschel15}. Our approach could also be used to investigate the orbital texture of exotic and stable hidden quantum states (\textit{e.g.} in 1T-TaS$2$~\cite{Stojchevska14}) which are not accessible using equilibrium tuning parameters. 

Moreover, applying stress to the material investigated here (1T-TiTe$_2$) has been predicted to drive strain-induced topological phase transitions~\cite{zhang_pressure-induced_2018} associated with a change of the orbital texture. Strain-dependent measurement of the \textit{i}LDAD could thus allow providing energy- and momentum-resolved view on the modification of orbital texture underlying topological phase transitions. This approach could also be used to investigate the role of orbital texture in the emergence of spontaneous electronic nematic ordering giving rise to the momentum-anisotropy of superconducting gap, for example in FeSe~\cite{rhodes_non-local_2021}.

One additional route that we envision is the ultrafast time-resolved investigation of dynamical
modification of orbital texture upon impulsive photoexcitation of solids. Angle-resolved photoemission spectroscopy in general, and more specifically the experimental setup used in this study, is compatible with time-resolved studies with femtosecond temporal resolution~\cite{Puppin19,Maklar20,Beaulieu2021_2}. Thus, combining this differential observable with pump-probe schemes could allow following in real-time modification of orbital texture during non-equilibrium topological phase transitions~\cite{vaswani20,Luo21}, photoinduced orbital order~\cite{grandi_fluctuation_2021-1} and coherent phonon excitation breaking fundamental symmetries of the materials~\cite{Sie19}. 


\section{Methods}\label{sec:methods}

\subsection{Angle-resolved photoemission spectroscopy}

The experimental apparatus features a table-top femtosecond XUV beamline coupled to a photoemission end-station. Briefly, a home-built optical parametric chirped-pulse amplifier (OPCPA) delivering 15 W (800 nm, 30 fs) at 500 kHz repetition rate~\cite{Puppin15} is used to drive high-order harmonic generation (HHG) by tightly focusing the second harmonic of the laser pulses (400 nm) onto a thin and dense Argon gas jet.  The nonperturbative nonlinear interaction between the laser pulses and the Argon atoms leads to the generation of a comb of odd harmonics of the driving laser, extending up to the 11th order. A single harmonic (7th order, 21.7 eV) is isolated by reflection on a focusing multilayer XUV mirror and propagation through a 400 nm thick Sn metallic filter. A photon flux of up to 2x10$^{11}$ photons/s at the sample position is obtained (110 meV FWHM). The bulk 1T-TiTe$_2$ samples (HQ Graphene) were cleaved at room temperature and base pressure of 5x10$^{-11}$ mbar, and handled by a 6-axis manipulator (SPECS GmbH). The photoemission data is acquired using a time-of-flight momentum microscope (METIS1000, SPECS GmbH). This detector allows for simultaneous detection of the full surface Brillouin zone, over an extended binding energy range, without the need to rearrange the sample geometry~\cite{Medjanik17}. For the data post-processing,  we use a recently developed open-source workflow~\cite{Xian19} to efficiently convert the raw single-event-based datasets into binned calibrated data hypervolumes of the desired dimension, including axes calibration and artifact corrections (including symmetry distortion corrections~\cite{Xian19_2}). The resulting 3D photoemission intensity data have the coordinates $I(E,k_x, k_y)$. 

\subsection{Density functional theory and tight-binding model}

We have performed DFT calculations for both the bulk and the monolayer 1T-TiTe$_2$ using the \textsc{Quantum Espresso} code~\cite{giannozzi_quantum_2009}. All calculations were performed using the PBE functional and norm-conserving pseudopotentials from the \textsc{PseudoDojo} project~\cite{van_setten_pseudodojo_2018}. Self-consistent calculations were converged on a $15\times 15$ Monkhorst-Pack grid for the Brillouin zone. For the monolayer, we constructed projective Wannier on the same grid with the \textsc{Wannier90} code~\cite{pizzi_wannier90_2020}.
We included the Ti-$d$ orbitals and the $p$ orbitals on each of the two Te atoms, yielding an 11-band TB model. This procedure yields the TB Hamiltonian $H^\mathrm{TB}_{jj^\prime}(\vec{k})$ and thus the coefficients defining the Bloch wavefunction in Eq.~\eqref{eq:blochwf}.
The real-space dependence of the Wannier functions is approximated by
\begin{align}
  \label{eq:pwfs}
  \phi_j(\vec{r}) = R_j(r)X_{\ell_j m_j}(\hat{\vec{r}}) \ ,
\end{align}
where $X_{\ell m}(\hat{\vec{r}})$ denotes the real spherical harmonics. Due to small weight of the $p$ orbitals we only construct the Ti $d$ orbitals explicitly; the radial dependence entering $R_j(r)$ is taken from the $d_{z^2}$ Wannier orbital and fixed for all $d$ orbitals. We use atomic units in what follows.

From the Wannier functions~\eqref{eq:pwfs} and the eigenvectors $C_{jn}(\vec{k})$ we construct the Bloch states via Eq.~\eqref{eq:blochwf},
which allows us to calculated the ARPES intensity according to Fermi's Golden rule
\begin{align}
\label{eq:fermigolden}
  I(\vec{k},E) &= \sum_n f_n(\vec{k}) | M_n(\vec{k},E)|^2 \delta(\varepsilon_n(\vec{k}) + \omega - E) \ , \\
  \label{eq:matrixelement}
  M_n(\vec{k},E) &= \langle \vec{k}, E| \vec{e}\cdot \hat{\vec{D}} | \psi_{\vec{k}n} \rangle \ .
\end{align}
Here, $f_n(\vec{k})$ is the occupation of band $n$ with energy $\varepsilon_n(\vec{k})$, $\omega$ is the photon energy, and $E$ denotes the energy of the final state $|\vec{k},E\rangle$. The band structure is shifted to reproduce the experimentally measured work function. The dipole operator is evaluate in the dipole gauge ($\hat{\vec{D}} = \vec{r}$), which is defined in terms of the Berry connection. We ignore nonlocal dipole contributions since photoemission at the M/M' pockets is predominantly originating from the Ti $d$ orbitals. The final states are approximated by plane waves. In this set-up, the dipole gauge has proven to be more accurate than the velocity gauge ($\hat{\vec{D}} = -i \nabla$) for a number of related systems~\cite{Schueler20,Beaulieu20_2,Day19}. Within this TB+PW model, the photoemission matrix element is then given by
\begin{align}
    \label{eq:mel_pw}
    M_n(\vec{k},E) = \sum_j C_{jn}(\vec{k})\int d \vec{r} e^{-i \vec{p}\cdot \vec{r}} \vec{e} \cdot \vec{r} \phi_j(\vec{r}) \ .
\end{align}
Here, $\vec{p} = (k_x,k_y,p_\perp)$ is the three-dimensional momentum of the photoelectron; the out-of-plane component $p_\perp$ is fixed by the kinetic energy $E = \vec{p}^2/2$. The orbital-resolved photoemission analysis in Section~\ref{subsec:tbarpes} is performed by including a subset of $j\in \{d_{z^2}, d_{xz}, d_{yz}, d_{x^2-y^2}, d_{xy}\}$ orbitals in Eq.~\eqref{eq:mel_pw}.

\subsection{One-step model of photoemission: KKR method}

The one-step model of photoemission is implemented in the spin-polarized relativistic Korringa-Kohn-Rostoker (SPR-KKR) scheme of the Munich bandstructure software package, based on Green’s function and multiple scattering spin-density matrix formalisms~\cite{ebert_calculating_2011,braun_correlation_2018}. Relativistic phenomena such as spin-orbit coupling (SOC) are treated by the Dirac equation. The local density approximation (LDA) has been chosen to approximate the exchange-correlation part of the potential. The bulk potential is converged within the atomic sphere approximation (ASA) geometry where the empty spheres were used. We have used a lattice constant of 3.779  \r{A} which corresponds to the interlayer Te-Te distance of 4.015 \r{A} . During the tuning procedure of the theoretical model, it was necessary to modify the Wigner-Seitz radius of individual atomic types to the following ratio: Ti = 1.03, Te= 1.08 and the vacuum type = 1. We used an angular momentum expansion up to a l$_{max}$=3. After the self-consistency was reached, the ARPES calculations have been performed and are based on the one-step model of photoemission in its spin density matrix formulation using the same geometry as used in the experiments. This theory accounts for effects induced by the light polarization, matrix-element effects, final state effects and surface effects. The final states are described by time-reversed low-energy electron diffraction (TRLEED) states. These states are calculated by multiple scattering formalism and it are considered to give the best single particle description of final states in photoemission. In particular, this approach goes far beyond free-electron-like approximation for the final state. 
Lifetime effects in the final states have been simulated via a constant imaginary part $V_{0f}$ = 1~eV in the inner potential and the lifetime of the initial state was simulated by an imaginary part of $V_{0i}$ =  0.01~eV. The photoemission signal factors in all the matrix element effects such as experimental geometry, photon energy, polarization state, and surface termination or final state effects.

\textbf{Acknowledgement} This work was funded by the Max Planck Society, the European Research Council (ERC) under the European Union’s Horizon 2020 research and innovation program (Grant No. ERC-2015-CoG-682843, H2020-FETOPEN-2018-2019-2020-01 (OPTOLogic - grant agreement No. 899794)), and the Deutsche Forschungsgemein-schaft (DFG, German Research Foundation) within the Emmy Noether program (Grant No. RE 3977/1), the Collaborative Research Center/Transregio 227  "Ultrafast Spin Dynamics" (project A09 and B07), the W\"urzburg-Dresden Cluster of Excellence on Complexity and Topology in Quantum Matter -- \textit{ct.qmat} (EXC 2147, project-id 39085490), and the Priority Program SPP 2244 (project No.~443366970).  J.S. and J.M. would like to thank CEDAMNF project financed by the Ministry of Education, Youth and Sports of Czech Republic, Project No. CZ.02.1.01/0.0/0.0/15.003/0000358. M.S. thanks the Alexander von Humboldt Foundation for its support with a Feodor Lynen scholarship. S.B. acknowledges financial support from the NSERC-Banting Postdoctoral Fellowships Program.\\

\textbf{Data availability} All experimental photoemission data used for the presented analysis will be publicly available on Zenodo open-access repository after acceptation of the manuscript. 

\textbf{Author contributions} S.B. acquired, analyzed and interpreted the experimental data. S.D., T.P., Ju.M. and A.N. participated in maintaining and running the experimental apparatus. M.S. performed and interpreted the tight-binding calculations. J.S. performed, J.S. and Ja.M. interpreted the KKR calculations with the help of F.R.. M.W., L.R. and R.E. were responsible for developing the infrastructures allowing these measurements. S.B. and M.S. wrote the first draft with inputs of Ja.M.. All authors contributed to the final version of the manuscript.


\begin{thebibliography}{10}

\bibitem{Damascelli04}
A.~Damascelli, Probing the electronic structure of complex systems by {ARPES}.
\newblock {\it Physica Scripta\/} {\bf T109}, 61 (2004).

\bibitem{Sobota21}
J.~A. Sobota, Y.~He, Z.-X. Shen, Angle-resolved photoemission studies of
  quantum materials.
\newblock {\it Rev. Mod. Phys.\/} {\bf 93}, 025006 (2021).

\bibitem{Hertz87}
H.~Hertz, Ueber einen einfluss des ultravioletten lichtes auf die electrische
  entladung.
\newblock {\it Annalen der Physik\/} {\bf 267}, 983-1000 (1887).

\bibitem{Einstein05}
A.~Einstein, Über einen die erzeugung und verwandlung des lichtes betreffenden
  heuristischen gesichtspunkt.
\newblock {\it Annalen der Physik\/} {\bf 322}, 132-148 (1905).

\bibitem{Hufner99}
S.~Hüfner, R.~Claessen, F.~Reinert, T.~Straub, V.~Strocov, P.~Steiner,
  Photoemission spectroscopy in metals: band structure-Fermi
  surface–spectral function.
\newblock {\it Journal of Electron Spectroscopy and Related Phenomena\/} {\bf
  100}, 191-213 (1999).

\bibitem{Bansil99}
A.~Bansil, M.~Lindroos, Importance of matrix elements in the ARPES spectra of
  BISCO.
\newblock {\it Phys. Rev. Lett.\/} {\bf 83}, 5154--5157 (1999).

\bibitem{Claessen92}
R.~Claessen, R.~O. Anderson, J.~W. Allen, C.~G. Olson, C.~Janowitz, W.~P.
  Ellis, S.~Harm, M.~Kalning, R.~Manzke, M.~Skibowski, Fermi-liquid line shapes
  measured by angle-resolved photoemission spectroscopy on 1T-TiTe$_2$.
\newblock {\it Phys. Rev. Lett.\/} {\bf 69}, 808--811 (1992).

\bibitem{Claessen96}
R.~Claessen, R.~O. Anderson, G.-H. Gweon, J.~W. Allen, W.~P. Ellis,
  C.~Janowitz, C.~G. Olson, Z.~X. Shen, V.~Eyert, M.~Skibowski, K.~Friemelt,
  E.~Bucher, S.~H\"ufner, Complete band-structure determination of the
  quasi-two-dimensional Fermi-liquid reference compound TiTe$_2$.
\newblock {\it Phys. Rev. B\/} {\bf 54}, 2453--2465 (1996).

\bibitem{Allen94}
P.~B. Allen, N.~Chetty, TiTe$_2$: Inconsistency between transport
  properties and photoemission results.
\newblock {\it Phys. Rev. B\/} {\bf 50}, 14855--14859 (1994).

\bibitem{perfetti01}
L.~Perfetti, C.~Rojas, A.~Reginelli, L.~Gavioli, H.~Berger, G.~Margaritondo,
  M.~Grioni, R.~Ga\'al, L.~Forr\'o, F.~Rullier~Albenque, High-resolution
  angle-resolved photoemission investigation of the quasiparticle scattering
  processes in a model Fermi liquid: 1T-TiTe$_2$.
\newblock {\it Phys. Rev. B\/} {\bf 64}, 115102 (2001).

\bibitem{nicolay06}
G.~Nicolay, B.~Eltner, S.~H\"ufner, F.~Reinert, U.~Probst, E.~Bucher,
  Importance of many-body effects to the spectral function of
  1T-TiTe$_2$.
\newblock {\it Phys. Rev. B\/} {\bf 73}, 045116 (2006).

\bibitem{Krasovskii07}
E.~E. Krasovskii, K.~Rossnagel, A.~Fedorov, W.~Schattke, L.~Kipp, Determination
  of the hole lifetime from photoemission: Ti $3d$ states in
  TiTe$_2$.
\newblock {\it Phys. Rev. Lett.\/} {\bf 98}, 217604 (2007).

\bibitem{Moser17}
S.~Moser, An experimentalist's guide to the matrix element in angle resolved
  photoemission.
\newblock {\it Journal of Electron Spectroscopy and Related Phenomena\/} {\bf
  214}, 29-52 (2017).

\bibitem{Day19}
R.~P. Day, B.~Zwartsenberg, I.~S. Elfimov, A.~Damascelli, Computational
  framework chinook for angle-resolved photoemission spectroscopy.
\newblock {\it npj Quantum Materials\/} {\bf 4}, 54 (2019).

\bibitem{Inosov07}
D.~S. Inosov, J.~Fink, A.~A. Kordyuk, S.~V. Borisenko, V.~B. Zabolotnyy,
  R.~Schuster, M.~Knupfer, B.~B\"uchner, R.~Follath, H.~A. D\"urr,
  W.~Eberhardt, V.~Hinkov, B.~Keimer, H.~Berger, Momentum and energy dependence
  of the anomalous high-energy dispersion in the electronic structure of high
  temperature superconductors.
\newblock {\it Phys. Rev. Lett.\/} {\bf 99}, 237002 (2007).

\bibitem{xian20}
R.~P. Xian, V.~Stimper, M.~Zacharias, S.~Dong, M.~Dendzik, S.~Beaulieu,
  B.~Schölkopf, M.~Wolf, L.~Rettig, C.~Carbogno, S.~Bauer, R.~Ernstorfer, A
  machine learning route between band mapping and band structure.
\newblock {\it 	arXiv:2005.10210 \/} (2020).

\bibitem{Wang12}
X.-P. Wang, P.~Richard, Y.-B. Huang, H.~Miao, L.~Cevey, N.~Xu, Y.-J. Sun,
  T.~Qian, Y.-M. Xu, M.~Shi, J.-P. Hu, X.~Dai, H.~Ding, Orbital characters
  determined from Fermi surface intensity patterns using angle-resolved
  photoemission spectroscopy.
\newblock {\it Phys. Rev. B\/} {\bf 85}, 214518 (2012).

\bibitem{Park12}
J.-H. Park, C.~H. Kim, J.-W. Rhim, J.~H. Han, Orbital Rashba effect and its
  detection by circular dichroism angle-resolved photoemission spectroscopy.
\newblock {\it Phys. Rev. B\/} {\bf 85}, 195401 (2012).

\bibitem{Sterzi18}
A.~Sterzi, G.~Manzoni, A.~Crepaldi, F.~Cilento, M.~Zacchigna, M.~Leclerc,
  P.~Bugnon, A.~Magrez, H.~Berger, L.~Petaccia, F.~Parmigiani, Probing band
  parity inversion in the topological insulator $\mathrm{GeBi_2Te_4}$ by linear
  dichroism in ARPES.
\newblock {\it Journal of Electron Spectroscopy and Related Phenomena\/} {\bf
  225}, 23 - 27 (2018).

\bibitem{Louat19}
A.~Louat, B.~Lenz, S.~Biermann, C.~Martins, F.~m.~c. Bertran, P.~Le~F\`evre,
  J.~E. Rault, F.~Bert, V.~Brouet, ARPES study of orbital character, symmetry
  breaking, and pseudogaps in doped and pure
  ${\mathrm{Sr}}_{2}{\mathrm{IrO}}_{4}$.
\newblock {\it Phys. Rev. B\/} {\bf 100}, 205135 (2019).

\bibitem{bentmann_profiling_2021}
H.~Bentmann, H.~Maa\ss{}, J.~Braun, C.~Seibel, K.~A. Kokh, O.~E. Tereshchenko,
  S.~Schreyeck, K.~Brunner, L.~W. Molenkamp, K.~Miyamoto, M.~Arita, K.~Shimada,
  T.~Okuda, J.~Kirschner, C.~Tusche, H.~Ebert, J.~Min\'{a}r, F.~Reinert,
  Profiling spin and orbital texture of a topological insulator in full
  momentum space.
\newblock {\it Phys. Rev. B\/} {\bf 103}, L161107 (2021).

\bibitem{Liu11}
Y.~Liu, G.~Bian, T.~Miller, T.-C. Chiang, Visualizing electronic chirality and
  Berry phases in graphene systems using photoemission with circularly
  polarized light.
\newblock {\it Phys. Rev. Lett.\/} {\bf 107}, 166803 (2011).

\bibitem{Bao21}
C.~Bao, H.~Zhang, T.~Zhang, X.~Wu, L.~Luo, S.~Zhou, Q.~Li, Y.~Hou, W.~Yao,
  L.~Liu, P.~Yu, J.~Li, W.~Duan, H.~Yao, Y.~Wang, S.~Zhou, Experimental
  evidence of chiral symmetry breaking in Kekul\'e-ordered graphene.
\newblock {\it Phys. Rev. Lett.\/} {\bf 126}, 206804 (2021).

\bibitem{Beaulieu20_2}
S.~Beaulieu, J.~Schusser, S.~Dong, M.~Sch\"uler, T.~Pincelli, M.~Dendzik,
  J.~Maklar, A.~Neef, H.~Ebert, K.~Hricovini, M.~Wolf, J.~Braun, L.~Rettig,
  J.~Min\'ar, R.~Ernstorfer, Revealing hidden orbital pseudospin texture with
  time-reversal dichroism in photoelectron angular distributions.
\newblock {\it Phys. Rev. Lett.\/} {\bf 125}, 216404 (2020).

\bibitem{Schuler21}
M.~Schüler, T.~Pincelli, S.~Dong, T.~P. Devereaux, M.~Wolf, L.~Rettig,
  R.~Ernstorfer, S.~Beaulieu, Bloch wavefunction reconstruction using
  multidimensional photoemission spectroscopy.
\newblock {\it 	arXiv:2103.17168 \/} (2021).

\bibitem{Cho18}
S.~Cho, J.-H. Park, J.~Hong, J.~Jung, B.~S. Kim, G.~Han, W.~Kyung, Y.~Kim,
  S.-K. Mo, J.~D. Denlinger, J.~H. Shim, J.~H. Han, C.~Kim, S.~R. Park,
  Experimental observation of hidden Berry curvature in inversion-symmetric
  bulk $2H\text{\ensuremath{-}}{\mathrm{WSe}}_{2}$.
\newblock {\it Phys. Rev. Lett.\/} {\bf 121}, 186401 (2018).

\bibitem{Schuler20}
M.~Sch\"{u}ler, U.~D. Giovannini, H.~H\"{u}bener, A.~Rubio, M.~A. Sentef,
  P.~Werner, Local {Berry} curvature signatures in dichroic angle-resolved
  photoelectron spectroscopy from two-dimensional materials.
\newblock {\it Science Advances\/} {\bf 6}, eaay2730 (2020).

\bibitem{Cho21}
S.~Cho, J.-H. Park, S.~Huh, J.~Hong, W.~Kyung, B.-G. Park, J.~D. Denlinger,
  J.~H. Shim, C.~Kim, S.~R. Park, Studying local Berry curvature in 2H-WSe2 by
  circular dichroism photoemission utilizing crystal mirror plane.
\newblock {\it Scientific Reports\/} {\bf 11}, 1684 (2021).

\bibitem{unzelmann_momentum-space_2021}
M.~\"{U}nzelmann, H.~Bentmann, T.~Figgemeier, P.~Eck, J.~N. Neu, B.~Geldiyev,
  F.~Diekmann, S.~Rohlf, J.~Buck, M.~Hoesch, M.~Kall\"{a}ne, K.~Rossnagel,
  R.~Thomale, T.~Siegrist, G.~Sangiovanni, D.~D. Sante, F.~Reinert,
  Momentum-space signatures of {Berry} flux monopoles in the {Weyl} semimetal
  {TaAs}.
\newblock {\it Nature Communications\/} {\bf 12}, 3650 (2021).

\bibitem{Terashima03}
K.~Terashima, T.~Sato, H.~Komatsu, T.~Takahashi, N.~Maeda, K.~Hayashi,
  Charge-density wave transition of
  $1T{\ensuremath{-}\mathrm{V}\mathrm{S}\mathrm{e}}_{2}$ studied by
  angle-resolved photoemission spectroscopy.
\newblock {\it Phys. Rev. B\/} {\bf 68}, 155108 (2003).

\bibitem{Watson19}
M.~D. Watson, O.~J. Clark, F.~Mazzola, I.~Markovi\ifmmode~\acute{c}\else
  \'{c}\fi{}, V.~Sunko, T.~K. Kim, K.~Rossnagel, P.~D.~C. King, Orbital- and
  ${k}_{z}$-selective hybridization of Se $4p$ and Ti $3d$ states in the charge
  density wave phase of ${\mathrm{TiSe}}_{2}$.
\newblock {\it Phys. Rev. Lett.\/} {\bf 122}, 076404 (2019).

\bibitem{Strokov12}
V.~N. Strocov, M.~Shi, M.~Kobayashi, C.~Monney, X.~Wang, J.~Krempasky,
  T.~Schmitt, L.~Patthey, H.~Berger, P.~Blaha, Three-dimensional electron realm
  in ${\mathrm{VSe}}_{2}$ by soft-x-ray photoelectron spectroscopy: Origin of
  charge-density waves.
\newblock {\it Phys. Rev. Lett.\/} {\bf 109}, 086401 (2012).

\bibitem{Chen17}
P.~Chen, W.~W. Pai, Y.-H. Chan, A.~Takayama, C.-Z. Xu, A.~Karn, S.~Hasegawa,
  M.~Y. Chou, S.-K. Mo, A.-V. Fedorov, T.-C. Chiang, Emergence of charge
  density waves and a pseudogap in single-layer TiTe$_2$.
\newblock {\it Nature Communications\/} {\bf 8}, 516 (2017).

\bibitem{Lin20}
M.-K. Lin, J.~A. Hlevyack, P.~Chen, R.-Y. Liu, S.-K. Mo, T.-C. Chiang, Charge
  instability in single-layer ${\mathrm{TiTe}}_{2}$ mediated by van der waals
  bonding to substrates.
\newblock {\it Phys. Rev. Lett.\/} {\bf 125}, 176405 (2020).

\bibitem{mor_ultrafast_2017}
S.~Mor, M.~Herzog, D.~Gole\v{z}, P.~Werner, M.~Eckstein, N.~Katayama,
  M.~Nohara, H.~Takagi, T.~Mizokawa, C.~Monney, J.~St\"{a}hler, Ultrafast
  {Electronic} {Band} {Gap} {Control} in an {Excitonic} {Insulator}.
\newblock {\it Phys. Rev. Lett.\/} {\bf 119}, 086401 (2017).

\bibitem{mazza_nature_2020}
G.~Mazza, M.~R\"{o}sner, L.~Windg\"{a}tter, S.~Latini, H.~H\"{u}bener, A.~J.
  Millis, A.~Rubio, A.~Georges, Nature of {Symmetry} {Breaking} at the
  {Excitonic} {Insulator} {Transition}: Ta$_2$NiSe$_5$.
\newblock {\it Phys. Rev. Lett.\/} {\bf 124}, 197601 (2020).

\bibitem{baldini_spontaneous_2020}
E.~Baldini, A.~Zong, D.~Choi, C.~Lee, M.~H. Michael, L.~Windgaetter, I.~I.
  Mazin, S.~Latini, D.~Azoury, B.~Lv, A.~Kogar, Y.~Wang, Y.~Lu, T.~Takayama,
  H.~Takagi, A.~J. Millis, A.~Rubio, E.~Demler, N.~Gedik, The spontaneous
  symmetry breaking in Ta$_2$NiSe$_5$ is structural in nature.
\newblock {\it arXiv:2007.02909\/}  (2020).

\bibitem{peng_observation_2021}
Y.~Peng, X.~Guo, Q.~Xiao, Q.~Li, J.~Strempfer, Y.~Choi, D.~Yan, H.~Luo,
  Y.~Huang, S.~Jia, O.~Janson, P.~Abbamonte, J.~v.~d. Brink, J.~van Wezel,
  Observation of orbital order in the {Van} der {Waals} material {1T}-{TiSe$_2$}.
\newblock {\it arXiv:2105.13195 \/} (2021).

\bibitem{Puppin19}
M.~Puppin, Y.~Deng, C.~W. Nicholson, J.~Feldl, N.~B.~M. Schröter, H.~Vita,
  P.~S. Kirchmann, C.~Monney, L.~Rettig, M.~Wolf, R.~Ernstorfer, Time- and
  angle-resolved photoemission spectroscopy of solids in the extreme
  ultraviolet at 500 kHz repetition rate.
\newblock {\it Review of Scientific Instruments\/} {\bf 90}, 023104 (2019).

\bibitem{Medjanik17}
K.~Medjanik, O.~Fedchenko, S.~Chernov, D.~Kutnyakhov, M.~Ellguth, A.~Oelsner,
  B.~Sch{\"o}nhense, T.~R.~F. Peixoto, P.~Lutz, C.-H. Min, F.~Reinert,
  S.~D{\"a}ster, Y.~Acremann, J.~Viefhaus, W.~Wurth, H.~J. Elmers,
  G.~Sch{\"o}nhense, Direct 3D mapping of the Fermi surface and Fermi velocity.
\newblock {\it Nature Materials\/} {\bf 16}, 615-621 (2017).

\bibitem{Maklar20}
J.~Maklar, S.~Dong, S.~Beaulieu, T.~Pincelli, M.~Dendzik, Y.~W. Windsor, R.~P.
  Xian, M.~Wolf, R.~Ernstorfer, L.~Rettig, A quantitative comparison of
  time-of-flight momentum microscopes and hemispherical analyzers for time- and
  angle-resolved photoemission spectroscopy experiments.
\newblock {\it Review of Scientific Instruments\/} {\bf 91}, 123112 (2020).

\bibitem{Seah79}
M.~P. Seah, W.~A. Dench, Quantitative electron spectroscopy of surfaces: A
  standard data base for electron inelastic mean free paths in solids.
\newblock {\it Surface and Interface Analysis\/} {\bf 1}, 2-11 (1979).

\bibitem{Riley14}
J.~M. Riley, F.~Mazzola, M.~Dendzik, M.~Michiardi, T.~Takayama, L.~Bawden,
  C.~Graner{\o}d, M.~Leandersson, T.~Balasubramanian, M.~Hoesch, T.~K. Kim,
  H.~Takagi, W.~Meevasana, P.~Hofmann, M.~.~S. Bahramy, J.~.~W. Wells, P.~.
  D.~C. King, Direct observation of spin-polarized bulk bands in an
  inversion-symmetric semiconductor.
\newblock {\it Nature Physics\/} {\bf 10}, 835-839 (2014).

\bibitem{Strokov06}
V.~N. Strocov, E.~E. Krasovskii, W.~Schattke, N.~Barrett, H.~Berger,
  D.~Schrupp, R.~Claessen, Three-dimensional band structure of layered
  ${\mathrm{TiTe}}_{2}$: Photoemission final-state effects.
\newblock {\it Phys. Rev. B\/} {\bf 74}, 195125 (2006).

\bibitem{Chernov15}
S.~Chernov, K.~Medjanik, C.~Tusche, D.~Kutnyakhov, S.~Nepijko, A.~Oelsner,
  J.~Braun, J.~Minár, S.~Borek, H.~Ebert, H.~Elmers, J.~Kirschner,
  G.~Schönhense, Anomalous d-like surface resonances on Mo(110) analyzed by
  time-of-flight momentum microscopy.
\newblock {\it Ultramicroscopy\/} {\bf 159}, 453 - 463 (2015).

\bibitem{Tusche16}
C.~Tusche, P.~Goslawski, D.~Kutnyakhov, M.~Ellguth, K.~Medjanik, H.~J. Elmers,
  S.~Chernov, R.~Wallauer, D.~Engel, A.~Jankowiak, G.~Schönhense, Multi-MHz
  time-of-flight electronic band structure imaging of graphene on Ir(111).
\newblock {\it Applied Physics Letters\/} {\bf 108}, 261602 (2016).

\bibitem{ebert_calculating_2011}
H.~Ebert, D.~K\"{o}dderitzsch, J.~Min\'{a}r, Calculating condensed matter
  properties using the {KKR}-{Green}'s function method—recent developments
  and applications.
\newblock {\it Rep. Prog. Phys.\/} {\bf 74}, 096501 (2011).

\bibitem{braun_correlation_2018}
J.~Braun, J.~Min\'{a}r, H.~Ebert, Correlation, temperature and disorder:
  {Recent} developments in the one-step description of angle-resolved
  photoemission.
\newblock {\it Phys. Rep.\/} {\bf 740}, 1--34 (2018).

\bibitem{zhou_theory_2020}
J.~S. Zhou, R.~Bianco, L.~Monacelli, I.~Errea, F.~Mauri, M.~Calandra, Theory of
  the thickness dependence of the charge density wave transition in 1
  {T}-{TiTe$_2$}.
\newblock {\it 2D Mater.\/} {\bf 7}, 045032 (2020).

\bibitem{smith_angular-resolved_1975}
N.~V. Smith, M.~M. Traum, Angular-resolved ultraviolet photoemission
  spectroscopy and its application to the layer compounds Ta${\mathrm{Se}}_{2}$
  and Ta${\mathrm{S}}_{2}$.
\newblock {\it Phys. Rev. B\/} {\bf 11}, 2087--2108 (1975).

\bibitem{traum_angular_1974}
M.~M. Traum, N.~V. Smith, F.~J. Di~Salvo, Angular dependence of photoemission
  and atomic orbitals in the layer compound
  $1T\ensuremath{-}\mathrm{Ta}{\mathrm{Se}}_{2}$.
\newblock {\it Phys. Rev. Lett.\/} {\bf 32}, 1241--1244 (1974).

\bibitem{Salvo76}
F.~J. Di~Salvo, D.~E. Moncton, J.~V. Waszczak, Electronic properties and
  superlattice formation in the semimetal ${\mathrm{TiSe}}_{2}$.
\newblock {\it Phys. Rev. B\/} {\bf 14}, 4321--4328 (1976).

\bibitem{Eaglesham86}
D.~J. Eaglesham, R.~L. Withers, D.~M. Bird, Charge-density-wave transitions in
  1T-VSe$_2$.
\newblock {\it Journal of Physics C: Solid State Physics\/} {\bf 19}, 359--367
  (1986).

\bibitem{ritschel15}
T.~Ritschel, J.~Trinckauf, K.~Koepernik, B.~B{\"u}chner, M.~v. Zimmermann,
  H.~Berger, Y.~I. Joe, P.~Abbamonte, J.~Geck, Orbital textures and charge
  density waves in transition metal dichalcogenides.
\newblock {\it Nature Physics\/} {\bf 11}, 328-331 (2015).

\bibitem{Stojchevska14}
L.~Stojchevska, I.~Vaskivskyi, T.~Mertelj, P.~Kusar, D.~Svetin, S.~Brazovskii,
  D.~Mihailovic, Ultrafast switching to a stable hidden quantum state in an
  electronic crystal.
\newblock {\it Science\/} {\bf 344}, 177--180 (2014).

\bibitem{zhang_pressure-induced_2018}
M.~Zhang, X.~Wang, A.~Rahman, Q.~Zeng, D.~Huang, R.~Dai, Z.~Wang, Z.~Zhang,
  Pressure-induced topological phase transitions and structural transition in
  {1T}-{TiTe$_2$} single crystal.
\newblock {\it Appl. Phys. Lett.\/} {\bf 112}, 041907 (2018).

\bibitem{rhodes_non-local_2021}
L.~C. Rhodes, J.~B\"{o}ker, M.~A. M\"{u}ller, M.~Eschrig, I.~M. Eremin,
  Non-local $d_{xy}$ nematicity and the missing electron pocket in {FeSe}.
\newblock {\it npj Quantum Materials\/} {\bf 6}, 1--14 (2021).

\bibitem{Beaulieu2021_2}
S.~Beaulieu, S.~Dong, N.~Tancogne-Dejean, M.~Dendzik, T.~Pincelli, J.~Maklar,
  R.~P. Xian, M.~A. Sentef, M.~Wolf, A.~Rubio, L.~Rettig, R.~Ernstorfer,
  Ultrafast dynamical lifshitz transition.
\newblock {\it Science Advances\/} {\bf 7}, eabd9275 (2021).

\bibitem{vaswani20}
C.~Vaswani, L.-L. Wang, D.~H. Mudiyanselage, Q.~Li, P.~M. Lozano, G.~D. Gu,
  D.~Cheng, B.~Song, L.~Luo, R.~H.~J. Kim, C.~Huang, Z.~Liu, M.~Mootz, I.~E.
  Perakis, Y.~Yao, K.~M. Ho, J.~Wang, Light-driven Raman coherence as a
  nonthermal route to ultrafast topology switching in a Dirac semimetal.
\newblock {\it Phys. Rev. X\/} {\bf 10}, 021013 (2020).

\bibitem{Luo21}
L.~Luo, D.~Cheng, B.~Song, L.-L. Wang, C.~Vaswani, P.~M. Lozano, G.~Gu,
  C.~Huang, R.~H.~J. Kim, Z.~Liu, J.-M. Park, Y.~Yao, K.~Ho, I.~E. Perakis,
  Q.~Li, J.~Wang, A light-induced phononic symmetry switch and giant
  dissipationless topological photocurrent in ZrTe$_5$.
\newblock {\it Nature Materials\/} {\bf 20}, 329--334 (2021).

\bibitem{grandi_fluctuation_2021-1}
F.~Grandi, M.~Eckstein, Fluctuation control of nonthermal orbital order.
\newblock {\it Phys. Rev. B\/} {\bf 103}, 245117 (2021).

\bibitem{Sie19}
E.~J. Sie, C.~M. Nyby, C.~D. Pemmaraju, S.~J. Park, X.~Shen, J.~Yang, M.~C.
  Hoffmann, B.~K. Ofori-Okai, R.~Li, A.~H. Reid, S.~Weathersby, E.~Mannebach,
  N.~Finney, D.~Rhodes, D.~Chenet, A.~Antony, L.~Balicas, J.~Hone, T.~P.
  Devereaux, T.~F. Heinz, X.~Wang, A.~M. Lindenberg, An ultrafast symmetry
  switch in a {Weyl} semimetal.
\newblock {\it Nature\/} {\bf 565}, 61--66 (2019).

\bibitem{Puppin15}
M.~Puppin, Y.~Deng, O.~Prochnow, J.~Ahrens, T.~Binhammer, U.~Morgner, M.~Krenz,
  M.~Wolf, R.~Ernstorfer, 500 kHz OPCPA delivering tunable sub-20 fs pulses
  with 15W average power based on an all-ytterbium laser.
\newblock {\it Opt. Express\/} {\bf 23}, 1491--1497 (2015).

\bibitem{Xian19}
R.~P. Xian, Y.~Acremann, S.~Y. Agustsson, M.~Dendzik, K.~B{\"u}hlmann,
  D.~Curcio, D.~Kutnyakhov, F.~Pressacco, M.~Heber, S.~Dong, T.~Pincelli,
  J.~Demsar, W.~Wurth, P.~Hofmann, M.~Wolf, M.~Scheidgen, L.~Rettig,
  R.~Ernstorfer, An open-source, end-to-end workflow for multidimensional
  photoemission spectroscopy.
\newblock {\it Scientific Data\/} {\bf 7}, 442 (2020).

\bibitem{Xian19_2}
R.~P. Xian, L.~Rettig, R.~Ernstorfer, Symmetry-guided nonrigid registration:
  The case for distortion correction in multidimensional photoemission
  spectroscopy.
\newblock {\it Ultramicroscopy\/} {\bf 202}, 133 - 139 (2019).

\bibitem{giannozzi_quantum_2009}
P.~Giannozzi, S.~Baroni, N.~Bonini, M.~Calandra, R.~Car, C.~Cavazzoni, {Davide
  Ceresoli}, G.~L. Chiarotti, M.~Cococcioni, I.~Dabo, A.~D. Corso, S.~d.
  Gironcoli, S.~Fabris, G.~Fratesi, R.~Gebauer, U.~Gerstmann, C.~Gougoussis,
  {Anton Kokalj}, M.~Lazzeri, L.~Martin-Samos, N.~Marzari, F.~Mauri,
  R.~Mazzarello, {Stefano Paolini}, A.~Pasquarello, L.~Paulatto, C.~Sbraccia,
  S.~Scandolo, G.~Sclauzero, A.~P. Seitsonen, A.~Smogunov, P.~Umari, R.~M.
  Wentzcovitch, {QUANTUM} {ESPRESSO}: a modular and open-source software
  project for quantum simulations of materials.
\newblock {\it J. Phys.: Condens. Matter\/} {\bf 21}, 395502 (2009).

\bibitem{van_setten_pseudodojo_2018}
M.~van Setten, M.~Giantomassi, E.~Bousquet, M.~Verstraete, D.~Hamann, X.~Gonze,
  G.-M. Rignanese, The {PseudoDojo}: {Training} and grading a 85 element
  optimized norm-conserving pseudopotential table.
\newblock {\it Comp. Phys. Commun.\/} {\bf 226}, 39--54 (2018).

\bibitem{pizzi_wannier90_2020}
G.~Pizzi, V.~Vitale, R.~Arita, S.~Bl\"{u}gel, F.~Freimuth, G.~G\'{e}ranton,
  M.~Gibertini, D.~Gresch, C.~Johnson, T.~Koretsune, J.~Iba\~{n}ez Azpiroz,
  H.~Lee, J.-M. Lihm, D.~Marchand, A.~Marrazzo, Y.~Mokrousov, J.~I. Mustafa,
  Y.~Nohara, Y.~Nomura, L.~Paulatto, S.~Ponc\'{e}, T.~Ponweiser, J.~Qiao,
  F.~Th\"{o}le, S.~S. Tsirkin, M.~Wierzbowska, N.~Marzari, D.~Vanderbilt,
  I.~Souza, A.~A. Mostofi, J.~R. Yates, Wannier90 as a community code: new
  features and applications.
\newblock {\it J. Phys.: Condens. Matter\/} {\bf 32}, 165902 (2020).

\bibitem{Schueler20}
M.~Sch\"{u}ler, U.~De~Giovannini, H.~H\"{u}bener, A.~Rubio, M.~A. Sentef, T.~P.
  Devereaux, P.~Werner, How {Circular} {Dichroism} in {Time}- and
  {Angle}-{Resolved} {Photoemission} {Can} {Be} {Used} to {Spectroscopically}
  {Detect} {Transient} {Topological} {States} in {Graphene}.
\newblock {\it Phys. Rev. X\/} {\bf 10}, 041013 (2020).


\end{thebibliography}

\section{Supplementary Information}

\subsection{Surface vs bulk sensitivity of the dichroic signal}

\begin{figure}
\centering\includegraphics[width=0.9\columnwidth]{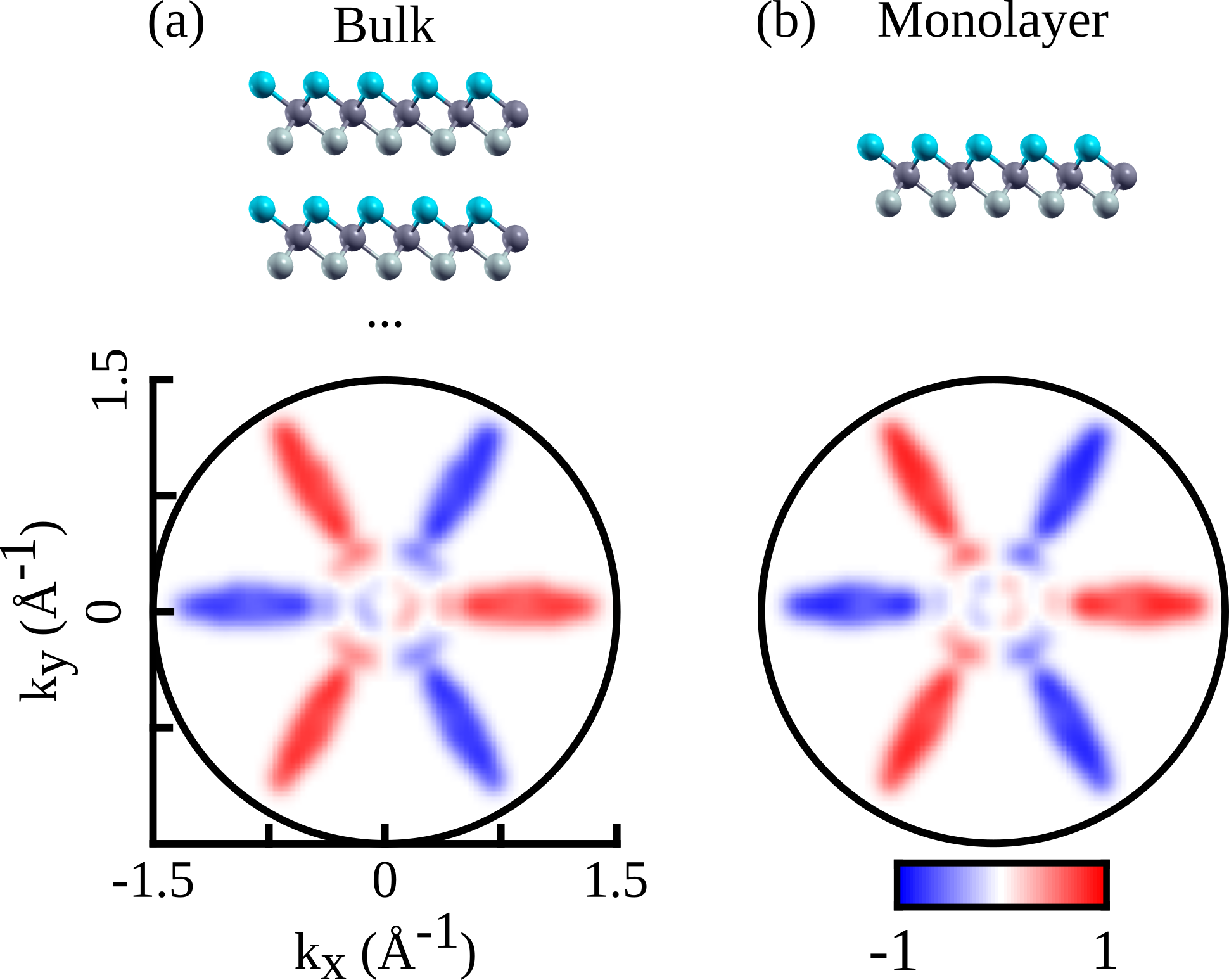}
\caption{\textbf{Layer-resolved \textit{i}LDAD calculated within the KKR framework}. \textit{i}LDAD for \textbf{(a)} bulk and \textbf{(b)} free-standing monolayer of 1T-TiTe$_2$ using 18.7~eV photon energy, and the same parameters as described in the methods.}
\label{fig:layer}
\end{figure}

In Fig.~\ref{fig:layer}, we compare \textit{i}LDAD for bulk and monolayer 1T-TiTe$_2$, calculated using the layer-resolved one-step model of photoemission described in the methods. The obtained \textit{i}LDAD for the bulk and monolayer cases are almost identical. Indeed, the features at both $\Gamma$ and M/M' points do not change when going from the monolayer case to the bulk limit. This striking observation gives us confidence that we can construct the TB model for a free-standing monolayer of 1T-TiTe$_2$ to get an intuitive picture about the orbital physics governing the emergence of the \textit{i}LDAD. 

\begin{figure}
\centering\includegraphics[width=0.9\columnwidth]{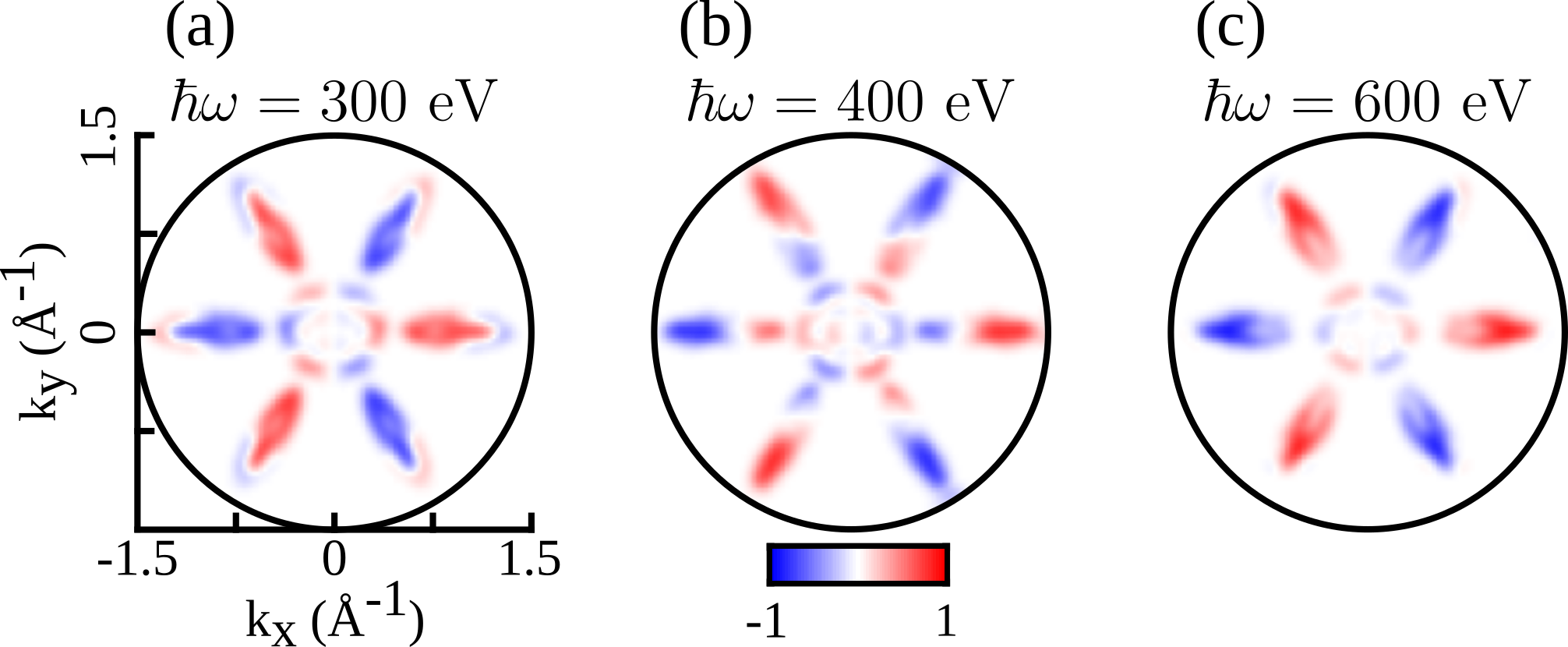}
\caption{\textbf{Soft-x-ray \textit{i}LDAD calculated within the KKR framework for different photon energy}. \textit{i}LDAD for bulk 1T-TiTe$_2$ has been calculated in the soft-x-ray regime at different photon energies \textbf{(a)} 300~eV, \textbf{(b)} 400~eV and \textbf{(c)} 600~eV. }
\label{fig:softxray}
\end{figure}

To probe the bulk \textit{i}LDAD spectrum, we performed corresponding calculations in the soft-x-ray regime. In Fig.~\ref{fig:softxray} we show photon-energy dependence of the \textit{i}LDAD for bulk 1T-TiTe$_2$, for 300~eV, 400~eV and 600~eV photon energies. The photon-energy-dependent trends from the XUV regime (see the main manuscript) are also recovered in the soft-x-ray regime. In particular, the dichroic features around the $\Gamma$ point reverse sign when increasing the photon energy. However, the observed \textit{i}LDAD at M and M' points, which is stable again variation of photon energy, further support the link between the in-plane orbital texture and \textit{i}LDAD. Furthermore, we point out that by increasing the photon energy to the soft-x-ray regime, TRLEED final states approach free-electron-like character, and the corresponding approximation in the TB model is justified for \textit{i}LDAD features around M and M' points. 

\subsection{Orbital weight analysis}

\begin{figure*}
\centering\includegraphics[width=0.95\textwidth]{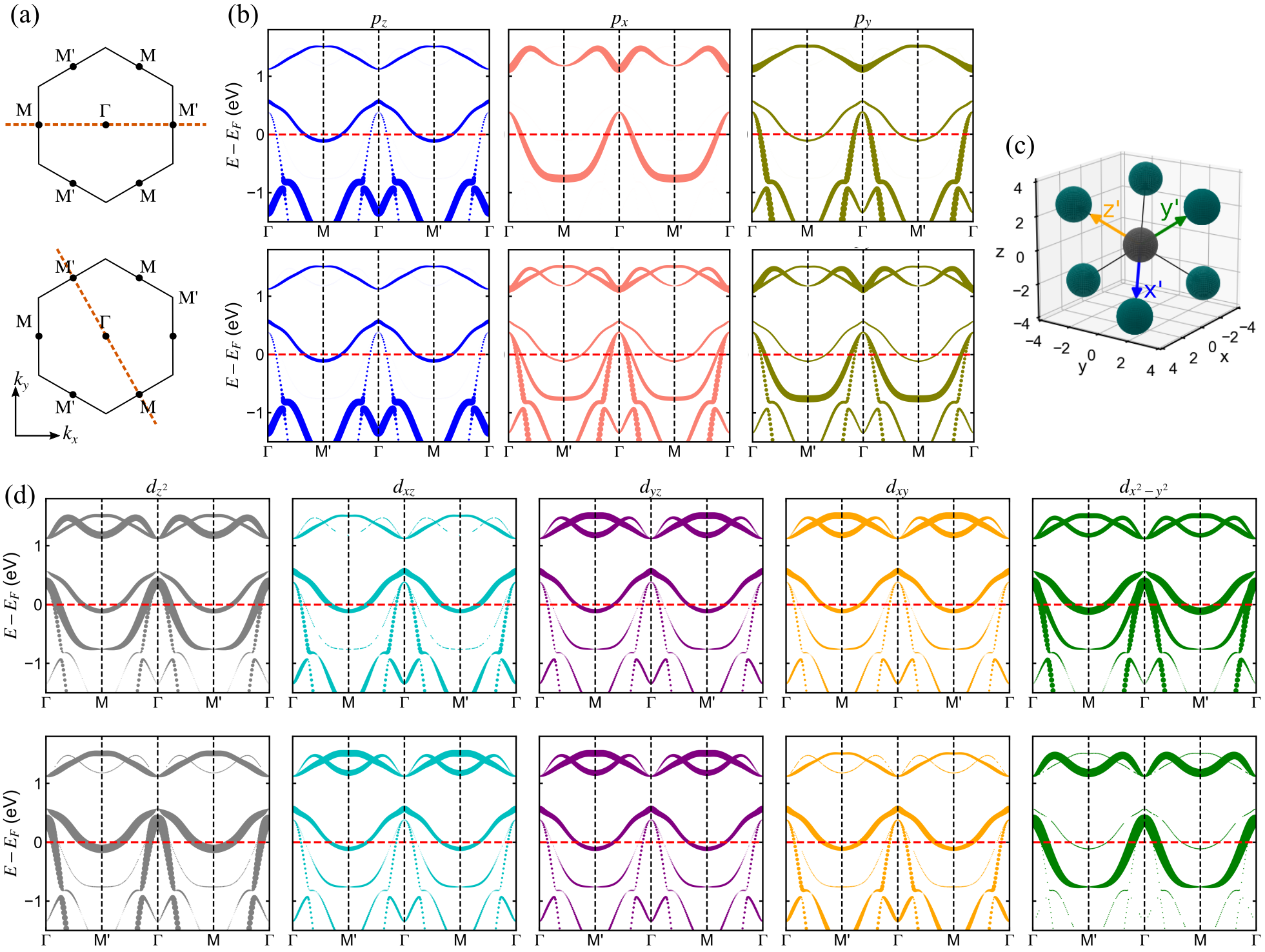}
\caption{\textbf{Orbital-projected band structure of monolayer 1T-TiTe$_2$}. \textbf{(a)} Sketch of the first Brillouin zone the paths in momentum space along which the band structures have been calculated. \textbf{(b)} Analogous to Fig.~(5) in the main text, the top (bottom) panels show the fat-band representation of the band structure along the path indicated in the top (bottom) in \textbf{(a)}. The thickness represents the summed weight from both Te atoms in the unit cell. The scale of the thickness is identical to Fig.~(5) in the main text. \textbf{(c)} Sketch of the modified geometry for defining the crystal-field coordinate system $(x^\prime, y^\prime, z^\prime)$. \textbf{(d)} Fat-band representation of the band structure (analogous to \textbf{(b)}) in the crystal-field basis.}
\label{fig:orbitals}
\end{figure*}

As explained in the main text and in the methods section, we have constructed a tight-binding (TB) Hamiltonian based on projective Wannier functions. This procedure yields the Hamiltonian $H_{j,j^\prime}(\vec{k})$, where $j, j^\prime$ run over the set of Te-$p$ and Ti-$d$ orbitals (11 orbitals in total). Diagonalizing $H_{j,j^\prime}(\vec{k})$ yields the eigenvalues $\varepsilon_n(\vec{k})$ and the associated eigenvectors $[\vec{C}_n(\vec{k})] = C_{jn}(\vec{k})$ (which enter Eq.~(4), Eq.~(6) and Eq.~(9) in the main text). 

The TB Hamiltonian is particularly useful for projecting onto specific orbitals. We define the orbital weight $w_{jn}(\vec{k}) = |C_{jn}(\vec{k})|^2$. Fig.~(5) in the main text shows the orbital weight for $j\in \{d_{z^2}, d_{xz}, d_{yz}\}$. The Te-$p$ orbitals play only a minor role at the M/M' pockets at the Fermi energy, which is confirmed by Fig.~\ref{fig:orbitals}(b). Contrasting the two paths in the Brillouin zone sketched in Fig/~\ref{fig:orbitals})(a), we not that the weight of the $p_z$ orbitals is constant, while the $p_x$ orbital has no contribution at M/M' for the path parallel to $k_x$. The weight of the $p_y$ orbital changes only slightly when considering the rotated path.

Refs.~\cite{traum_angular_1974,smith_angular-resolved_1975} attempted to explain the observed anisotropies in ARPES from the related compounds 1T-TaSe$_2$ and 1T-TaS$_2$ to a crystal-field splitting effect. We used the experimental geometry of the bulk system for all calculations, defined by the lattice constant $a = 3.75$~\AA{} and the vertical distance between the Te layers of $d_\mathrm{Te}=3.367$~\AA{}. In this geometry the crystal axes defined by the Ti-Te bonds, are non-orthogonal. Hence, the angular momentum is quenched, and the Ti-$d$ orbitals in the lab frame would be expressed by orbitals with different total angular momentum in the crystal-field basis. However, for a qualitative picture from the crystal-field point of view, we define the crystal-field axes in a modified geometry where bond directions are orthogonal (see Fig.~\ref{fig:orbitals}(c)). This is achieved by reducing the Te layer distance to $d^\prime_\mathrm{Te}=3.054$~\AA{}. The resulting ligand configuration is identical to the octahedral complex; the $d$ orbitals split into the $e_g$ orbitals ($d_{z^2}, d_{x^2-y^2}$) and $t_{2g}$ orbitals ($d_{xy}, d_{xz}, d_{yz}$) in the new coordinate system $(x^\prime, y^\prime, z^\prime)$.

Inspecting the orbital weight in the crystal-field basis of Ti-$d$ orbitals (Fig.~\ref{fig:orbitals}(d)) we find a qualitative difference between $e_g$ and $t_{2g}$ orbitals. The weight of the $t_{2g}$ at the M/M' pockets is almost unchanged when rotating the path in momentum space by $60^\circ$, while the $e_g$ orbitals are strongly affected. For the path parallel to the $k_x$ direction (upper panels in Fig.~\ref{fig:orbitals}(d)), the $d_{x^2-y^2}$ orbital dominates, while for the rotated path the $d_{x^2-y^2}$ transfers most of its weight to $d_{z^2}$. This analogous to the exchange of orbital weight between $d_{xz}$ and $d_{yz}$ orbitals in the lab frame discussed in the main text. 
The analysis in the crystal-field basis provides a complementary but ultimately not simpler picture, as the Bloch wavefunction contains notable contributions from both the $e_g$ and $t_{2g}$ orbitals.

\end{document}